\pdfoutput=1
\RequirePackage{ifpdf}
\ifpdf 
\documentclass[pdftex]{sigma}
\else
\documentclass{sigma}
\fi

\usepackage{mathtools}
\usepackage{xparse}

\DeclareMathOperator{\reg}{reg}
\DeclareMathOperator*{\res}{Res}
\DeclareMathOperator{\tr}{tr}
\newcommand{\Cset}{\mathbb{C}}
\newcommand{\Rset}{\mathbb{R}}
\newcommand{\la}{\lambda}
\newcommand{\ho}{\tilde}
\newcommand{\Ho}{\widetilde}
\NewDocumentCommand{\Fmat}{s}{\IfBooleanTF{#1}{\Ho{\mathcal{F}}}{\mathcal{F}}}
\newcommand{\Sla}{\mathcal{M}}
\newcommand{\FW}{\mathcal{W}}
\NewDocumentCommand{\Gau}{s}{\IfBooleanTF{#1}{\Ho{\mathcal{N}}}{\mathcal{N}}}
\newcommand{\Id}{\mathcal{I}}
\NewDocumentCommand{\aux}{s}{\IfBooleanTF{#1}{\ho{\mathfrak{a}}}{\mathfrak{a}}}
\newcommand{\Aux}{\mathcal{A}}
\NewDocumentCommand{\den}{s}{\IfBooleanTF{#1}{\ho{\rho}}{\rho}}
\NewDocumentCommand{\Den}{s}{\IfBooleanTF{#1}{\Ho{\mathcal{R}}}{\mathcal{R}}}
\newcommand{\Hmat}{\Ho{\mathcal{H}}}
\newcommand{\rlx}{r}
\newcommand{\rcx}{rc+}
\newcommand{\clpx}{c}
\newcommand{\wdpx}{w}

\newcommand{\rl}{\gamma}
\newcommand{\hle}{\theta}
\NewDocumentCommand{\clp}{s O{}}{\IfBooleanTF{#1}{\bar{\xi}}{\xi}}\NewDocumentCommand{\wdp}{s}{\IfBooleanTF{#1}{\bar{\omega}}{\omega}}\newcommand{\sdv}{\delta}
\numberwithin{equation}{section}

\newcommand{\bra}[1]{\langle #1\,|}
\newcommand{\ket}[1]{|\,#1\rangle}
\newcommand{\braket}[2]{\ensuremath{\langle #1 \mid #2 \rangle }}
\newcommand{\set}[1]{\lbrace#1\rbrace}

\def\pl{\prod\limits}

\begin{document}
\allowdisplaybreaks

\renewcommand{\thefootnote}{}

\newcommand{\arXivNumber}{2105.10244}

\renewcommand{\PaperNumber}{112}

\FirstPageHeading

\ShortArticleName{Form Factors of the Heisenberg Spin Chain in the Thermodynamic Limit}

\ArticleName{Form Factors of the Heisenberg Spin Chain\\ in the Thermodynamic Limit:\\ Dealing with Complex Bethe Roots\footnote{This paper is a~contribution to the Special Issue on Mathematics of Integrable Systems: Classical and Quantum in honor of Leon Takhtajan.

~~\,The full collection is available at \href{https://www.emis.de/journals/SIGMA/Takhtajan.html}{https://www.emis.de/journals/SIGMA/Takhtajan.html}}}

\Author{Nikolai KITANINE and Giridhar KULKARNI}

\AuthorNameForHeading{N.~Kitanine and G.~Kulkarni}

\Address{Institut de Math\'ematiques de Bourgogne, UMR 5584, CNRS,\\ Universit\'e Bourgogne Franche-Comt\'e, F-21000 Dijon, France}
\Email{\href{mailto:nikolai.kitanine@u-bourgogne.fr}{nikolai.kitanine@u-bourgogne.fr}, \href{mailto:giridhar.kulkarni@u-bourgogne.fr}{giridhar.kulkarni@u-bourgogne.fr}}

\ArticleDates{Received May 28, 2021, in final form December 17, 2021; Published online December 25, 2021}

\Abstract{In this article we study the thermodynamic limit of the form factors of the $XXX$ Heisenberg spin chain using the algebraic Bethe ansatz approach. Our main goal is to express the form factors for the low-lying excited states as determinants of matrices that remain finite dimensional in the thermodynamic limit. We show how to treat all types of the complex roots of the Bethe equations within this framework. In particular we demonstrate that the Gaudin determinant for the higher level Bethe equations arises naturally from the algebraic Bethe ansatz.}

\Keywords{spin chains; form factors; correlation functions; algebraic Bethe ansatz}

\Classification{81U15; 81U40; 45F05}

\renewcommand{\thefootnote}{\arabic{footnote}}
\setcounter{footnote}{0}

\begin{flushright}
\begin{minipage}{58mm}
\it To L.A.~Takhtajan on the occasion\\
 of his 70$\,{}^{th}$ birthday
\end{minipage}
\end{flushright}

\section{Introduction}
The computation of time dependent correlation functions for the spin chains remains one of the most important and challenging problems in the field of quantum integrability. Recent numerical computation of correlation functions based on the form factor analysis leading to predictions for experimentally measurable quantities \cite{MouEKCSR13} have greatly enhanced the interest in this problem.
	
Typically we want to compute a ground state mean value of products of time-dependent local spin operators. The most straightforward way to proceed with this computation is to introduce a complete set of eigenstates of the Hamiltonian between the spin operators. For the most interesting example of the two-point time dependent function one obtains a sum over all the eigenstates of the spin chain
\begin{gather*}
\langle\sigma_{m+1}^a(t)\sigma_1^a\rangle=\sum\limits_{\Psi_{\mathrm{ex}}}\exp(-{\rm i}\Delta E_{\mathrm{ex}}t+{\rm i}\Delta P_{\mathrm{ex}}m) \big|\langle\Psi_g|\sigma_1^a |\Psi_{\mathrm{ex}}\rangle\big|^2,\qquad a=x,y,z.
\end{gather*}
Here $\Delta E_{\mathrm{ex}}$ and $\Delta P_{\mathrm{ex}}$ are energy and momentum differences between the corresponding excited state $\ket{\Psi_{\mathrm{ex}}}$ and the ground state $\ket{\Psi_{g}}$ while the remaining coefficient is the {\it form factor}.

One can argue that the main contribution in the thermodynamic limit to the two point function can be obtained taking into account only the low-lying excitations close to the ground state. Thus the problem of the computation of the correlation functions can be reduced to the description of the low-lying excitations, computation of the corresponding energies and momenta and finally computation of the form factors.

This scheme evidently can be applied to any quantum model but for the spin chains due to the integrability all the quantities in this series can be computed explicitly. The first description of the eigenstates was obtained by H.~Bethe \cite{Bet31}, however it is now well established that for the computation of form factors and correlation functions the most efficient description of the eigenstates is the {\it algebraic Bethe ansatz} introduced by Faddeev, Sklyanin and Takhtajan~\cite{FadST79}. It is also the work of Faddeev and Takhtajan that permitted to describe systematically the low-lying excitations \cite{FadT84, FadT81a} in terms of holes (or spinons) and to compute the corresponding energies and momenta. It was shown that the low-lying excitations can be defined in terms of even number of real valued rapidities (hole positions). We would like to emphasise that this number of holes remains finite in the thermodynamic limit.

Thus the remaining problem is a systematic computation of form factor for the quantum integrable systems. This is a long-standing problem and numerous approaches were used to compute these quantities \cite{JimM95L,KitMT99, Smi92L}. For the spin chains the two main approaches to the form factors and correlation functions are based either on the $q$-vertex operator algebra \cite{JImMMN92,JimM95L} or on the algebraic Bethe ansatz \cite{KitMT99,KitMT00}. Interestingly while the two approaches lead to the same multiple integral representations for the correlation functions the form factor analysis remained very different. The $q$-vertex operator approach leads to simple expressions for the low-lying exited states for the spin chains without external magnetic field in particular for the~2 and~4 spinon states for the $XXX$ chain \cite{AbaBS97,BouCK96,CauH06} and for the $XXZ$ chain in the massless regime~\cite{CauKSW12}. The algebraic Bethe ansatz approach permitted to deal with the non-zero magnetic field case leading to Fredholm determinant representations for the form factors \cite{KitKMST09c,KitKMST11a,Koz17}. These representations turned out to be extremely useful for the asymptotic analysis of the correlation functions \cite{KitKMST11b,KitKMST12,Koz19}. However this approach did not permit to reproduce the results for the form factors without an external magnetic field.

In this paper we propose an approach based on the algebraic Bethe ansatz solution for the $XXX$ spin chain without external magnetic field. The determinant representation we use as the starting point for this paper were obtained in \cite{KitMT99} using the Slavnov determinant representation for the scalar products of off-shell and on-shell Bethe vectors \cite{Sla89}, Gaudin--Korepin formula for the norms of Bethe vectors \cite{Gau83L,GauMcCW81,KirK88,Kor82} and the solution of the quantum inverse problem~\mbox{\cite{KitMT99,MaiT00}}. The form factors are thus represented as ratio of two determinants of $M/2\times M/2$ matrices, where $M$ is the length of the chain. In the thermodynamic limit the matrices become infinite and their analysis becomes rather complicated. In particular it can be shown that both determinants (in the numerator and denominator) are divergent in the thermodynamic limit if we consider the spin chains without external magnetic field.

 The main goal of this paper is to analyse these determinants in the thermodynamic limit for the low-lying excited states with finite number of holes and to take into account all the types of complex roots. It turns out that contrary to the energy and momentum completely defined by the position of holes the values of form factors depend also on the position of complex roots of Bethe equations. The most systematic description of these complex roots for the low-lying excited states without any assumptions (such as the string hypothesis) was introduced by Destri and Lowenstein in \cite{DesL82} and further developed in \cite{BabVV83, Woy82}. In particular it was shown that the positions of complex roots are determined by the position of holes through a set of equations very similar to the Bethe equations for a inhomogeneous spin chain with a number of sites equal to the number of holes. These equations are called higher level Bethe equations. While energy and momentum do not depend on the position of complex roots to compute the form factors the solutions of these higher level Bethe equations should be taken into account.

 To proceed with computations of form factors we will use the technique introduced in our previous paper \cite{KitK19}, where the simplest case of the two-spinon form factors is considered. We~have shown that the ratio of two determinants can be computed for a long but finite lattice and leads to a simple Cauchy determinant which has easily computable thermodynamic limit. However this Cauchy determinant for a generic excitation is ``perturbed'' by a finite rank matrix depending on the complex roots. We show in this paper that the form factors can be represented as determinants of {\it finite dimensional matrices}. Moreover the main result of this paper is that the form factors for the $XXX$ Heisenberg spin chain in the thermodynamic limit can be written as a ratio of determinants of finite matrices that involves in it a Gaudin matrix for the higher level Bethe equations. We would like to mention that similar Gaudin matrix appears in the framework of a different approach in \cite{DugGKS15}.

 The approach used in this paper turns out to be rather efficient for the computation of form factors. In particular it was recently used for the computation of thermal form factors \cite{BabGKS21}.

 Finally it is important to point out that similar finite size determinant representations were obtained from the BJMST fermionic approach \cite{JimMS11}. It would be extremely interesting to better understand the relations between these two methods and two types of determinant representations.

The paper is organised as follows: in the next section we give a technical introduction to the algebraic Bethe ansatz, Destri--Lowenstein description of the low-lying excitation and determinant representations for the form factors. In the Section~\ref{sec:gauex} we introduce our method of computation of ratio of two determinants (here we call it Gaudin extraction) and get the modi\-fied Cauchy determinant representation for the form factors. Finally in the last section we obtain the final result and show how the Gaudin matrix of the higher level Bethe equations appears in this framework.

\section{Algebraic Bethe ansatz and the form factors}
We consider in this paper the isotropic $XXX$ Heisenberg spin chain without an external magnetic field
\begin{gather}
H_{XXX}=\sum_{m=1}^M\big(\sigma_m^x\sigma_{m+1}^x+\sigma_m^y\sigma_{m+1}^y+\sigma_m^z\sigma_{m+1}^z-1\big).
\label{Hamiltonian}
\end{gather}
with periodic boundary conditions $\sigma^a_{M+1}=\sigma_1^a$. The main goal of this paper is the computation of form factors
\begin{gather*}
\big|F^z(\Psi_{\mathrm{ex}})\big|^2=\big|\bra{\Psi_g}\sigma_1^z\ket{\Psi_{\mathrm{ex}}}\big|^2,
\end{gather*}
where $\ket{\Psi_g}$ is the ground state of the Hamiltonian (\ref{Hamiltonian}) and $\ket{\Psi_{\mathrm{ex}}}$ is a low-lying excited state. Note that the computation of the form factors for other spin operators ($\sigma^x$ or $\sigma^y$) is equivalent as we are dealing with an isotropic model.
In this section we will remind the construction of the eigenstates of the $XXX$ Hamiltonian in the algebraic Bethe ansatz framework~\cite{FadST79}, determinant formulas for the form factors for a finite chain \cite{KitMT99} and a description of the low-lying excitations following the Destri--Lowenstein approach~\cite{DesL82}.

\subsection{Algebraic Bethe ansatz}
In this subsection we give a brief reminder of the algebraic Bethe ansatz construction of the eigenstates. The central element of this construction is the rational $R$-matrix
acting in a tensor product $V_1\otimes V_2$ of 2 two-dimensional spaces $V_j\simeq\mathbb{C}^2$
\begin{gather*}
R_{12}(\la)=\frac 1{\la+{\rm i}}(\la I_{12}+{\rm i} P_{12}),
\end{gather*}
where $P_{12}$ is the permutation operator. It is the well-known rational solution the Yang--Baxter equation. For the $XXX$ chain it can be used as the local $L$-operator acting in the tensor product of the auxiliary space $V_0$ and a local quantum space $V_m$. Then the monodromy matrix can be defined as an ordered product of local $L$-operators
\begin{gather*}
T_0(\la)=R_{0M}\bigg(\la-\frac{\rm i}{2}\bigg)\cdots R_{01}\bigg(\la-\frac{\rm i}{2}\bigg)=\begin{pmatrix} A(\la) & B(\la)\\ C(\la) & D(\la)\end{pmatrix}_0\!.
\end{gather*}
The operators $A$, $B$, $C$ and $D$ act in the same space as the Hamiltonian of the $XXX$ chain and their commutation relations follows from the $RTT$ relation for the monodromy matrices
\begin{gather*}
R_{ab}(\la-\mu)T_a(\la)T_b(\mu)=T_b(\mu)T_a(\la)R_{ab}(\la-\mu).
\end{gather*}
In particular, this relation means that the transfer matrices
\begin{gather*}
\mathcal{T}(\la)=\tr_0 T_0(\la)=A(\la)+D(\la),
\end{gather*}
commute for different values of the spectral parameter
\begin{gather*}
[\mathcal{T}(\la),\mathcal{T}(\mu)]=0.
\end{gather*}
It can be easily shown that the Hamiltonian can be obtained as the logarithmic derivative of the transfer matrix
\begin{gather*}
H_{XXX}= 2{\rm i}\frac {\rm d}{{\rm d}\la}\log(\mathcal{T}(\la))\bigg|_{\la=\frac{\rm i}{2}},
\end{gather*}
and it means that the transfer matrix generates a family of conserved charges.

The eigenstates of the transfer matrix and thus of the Hamiltonian can be constructed using the operators $B$ as creation operators. We start from the ferromagnetic state with all the spins up denoted $\ket{0}$,
\begin{gather*}
C(\la)\ket{0}=0, \qquad
A(\la)\ket{0}=\ket{0}, \qquad
D(\la)\ket{0}=\bigg(\frac{\la-\frac{\rm i}{2}}{\la+\frac{\rm i}{2}}\bigg)^M\ket{0}, \qquad \forall\la\in\mathbb{C}.
\end{gather*}
Then a state constructed by action of operators $B(\la)$ on this ferromagnetic state
\begin{gather*}
\ket{\Psi(\la_1,\dots,\la_N)}=B(\la_1)\cdots B(\la_N)\ket{0},
\end{gather*}
will be called a Bethe state or more precisely an {\it off-shell Bethe state} for generic parameters~$\la_j$. It is well defined as operators $B$ commute for different values of the spectral parameter. A~Bethe state is an eigenstate of the transfer matrix (and thus of the Hamiltonian) if the spectral parameters satisfy the Bethe equations
\begin{gather}
\bigg(\frac{\la_j-\frac{\rm i}{2}}{\la_j+\frac{\rm i}{2}}\bigg)^M\prod_{k=1}^N\frac{\la_j-\la_k+{\rm i}}{\la_j-\la_k-{\rm i}}=-1
.
\label{Bethe_eq}
\end{gather}
If these equations are satisfied we will refer to the corresponding state as an {\it on-shell Bethe state}.
To simplify the notations for any on-shell Bethe state in the following we define the corresponding Baxter polynomial
\begin{gather}
q(\la)=\prod_{j=1}^N(\la-\la_j),
\label{baxq}
\end{gather}
and exponential counting function
\begin{gather}
\mathfrak{a}(\la)=\bigg(\frac{\la-\frac{\rm i}{2}}{\la+\frac{\rm i}{2}}\bigg)^M\frac{q(\la+{\rm i})}{q(\la-{\rm i})}.
\label{exp_count}
\end{gather}
Using these notations the Bethe equations (\ref{Bethe_eq}) can be written as
\begin{gather*}
\mathfrak{a}(\la_j)+1=0,\qquad
j=1,\dots, N.
\end{gather*}
The corresponding eigenvalue of the transfer matrix $\mathcal{T}(\mu)$ can be written as
\begin{gather}
\tau(\mu|\{\la\})= (\mathfrak{a}(\mu)+1)\frac{q(\mu-{\rm i})}{q(\mu)}.
\label{t-eigen}
\end{gather}
The dual Bethe vectors can be constructed similarly using operators $C(\la)$
\begin{gather*}
\bra{\Psi(\la_1,\dots,\la_N)}=\bra{0}C(\la_1)\cdots C(\la_N),
\end{gather*}
once again they are eigenvectors of the transfer matrix with eigenvalue given by (\ref{t-eigen}) if the Bethe equations (\ref{Bethe_eq}) are satisfied.

Finally we would like to mention that the isotropic $XXX$ model has an additional ${\rm SU}(2)$ symmetry.
Every on-shell Bethe vector is a highest weight vector
\begin{gather*}
S_+B(\la_1)\cdots B(\la_N)\ket{0}=0, \qquad S_+=\sum_{m=1}^M\sigma^+_m,
\end{gather*}
and gives rise to a multiplet of the transfer matrix eigenvectors with the same eigenvalue
\begin{gather}
\ket{\Psi_\ell(\la_1,\dots,\la_N)}=S_-^\ell \ket{\Psi(\la_1,\dots,\la_N)}, \qquad
\ell=0,1,\dots, M-2N.
\label{mult_xxx}
\end{gather}
Note that the action of operators $S_-$ can be considered as adding of $\ell $ quasiparticles with infinite rapidities as
\begin{gather*}
S_-=\lim_{\la\rightarrow\infty} \la B(\la).
\end{gather*}

\subsection{Ground state and excitations in the Destri--Lowenstein picture}
The ground state of the antiferromagnetic spin chains has a non-trivial form. It was first identified in \cite{YanY66}, where it was proved that the ground state is given by a Bethe vector with $\frac{M}{2}$ real roots of the Bethe equations.
In the thermodynamic limit, it forms a dense distribution of real Bethe roots and the corresponding density function $\den_g(\la)$ satisfies the integral equation
\begin{gather*}
 \den_g(\la)+\int_\Rset K_1(\la-\nu)\den_g(\nu){\rm d}\nu
 = K_{\frac{1}{2}}(\la),\qquad
 K_a(\la)=\frac{a}{\pi}\frac{1}{\la^2+a^2}.
\end{gather*}
This is known as the Lieb integral equation, its kernel will be often denoted simply as $K(\la)\allowbreak\equiv K_1(\la)$.
The solution of the Lieb integral equation can be obtained using the Fourier transform
\begin{gather*}
 \den_g(\la)=\frac{1}{2\cosh\pi\la} .
\end{gather*}
The density of Bethe roots for the $XXX$ model is supported over the entire real line, however it diminishes exponentially at infinity. We will use this observation to distinguish, in a rough sense, the bulk of distribution from its tails.

The excited states over antiferromagnetic ground state are generated by {\it Faddeev--Takhtajan spinons} \cite{FadT81}. 
In Bethe ansatz framework, it corresponds to the holes $\set{\hle_a}_{a\leq n_h}$ in the Fermi distribution of the real roots.
The density function for real roots of such an excited state therefore admits the decomposition
\begin{gather*}
 \den_e(\la)=\den_g(\la)+\frac{1}{M}\sum_{a=1}^{n_h}\den_h(\la-\hle_a), \qquad
 \den_h(\la)+\int_\Rset K(\la-\nu)\den_h(\nu){\rm d}\nu=K(\la) .
\end{gather*}
The spinon excitations obey the fractional statistical law as the number of holes is always an even number \cite{FadT81, FadT84}.
Moreover, one can also show that energy and momentum of spinons is given by the following expression in terms of the hole parameters
\begin{gather}
 E_e = E_g + \sum_{a=1}^{n_h}\frac{\pi}{2\cosh\pi\hle_a},
 \label{ene_eigen}
 \\
 P_e = P_g + \sum_{a=1}^{n_h}\arctan\sinh\pi\hle_a -\frac{\pi}{2}.
 \label{mom_eigen}
\end{gather}

In the thermodynamic limit, the complex roots are expected to form particular configurations. The string hypothesis gives a hierarchical organisation of Bethe roots into the string complexes of different lengths that can be arbitrarily large. Despite the utility of string hypothesis, its claim has been widely disputed. It has been known that strings of length greater than two do not form in the thermodynamic limit for excitations close to the antiferromagnetic ground state.
Due to this reason, we will not rely on strong assumptions of string hypothesis. We will use instead the Destri--Lowenstein picture \cite{DesL82}, where all non-real complex roots can be organised into close-pair $\big(|\operatorname{Im}\la|<\frac{1}{2}\big)$ or wide-pair $\big(|\operatorname{Im}\la|>\frac{1}{2}\big)$ categories.
In this paper we will use notations $(\clp^\pm)$ and $(\wdp^+, \wdp*^-)$ for close-pairs and wide-pairs respectively, where
\begin{gather*}
 z^\pm = z \pm \frac{\rm i}{2}.
\end{gather*}
In the thermodynamic limit, close-pairs condense to form 2-string or quartet configurations. Hence in the Destri--Lowenstein picture Bethe roots of an excited state are partitioned into
\begin{alignat}{3}
 & \text{real roots:}\quad&&
 \set{\rl_j}_{j=1}^{n_r},&
 \label{DL_rl}
 \\
 & \text{2-strings:}\quad &&
 \set{\clp^+_a+{\rm i}\sdv_a,\clp^-_a-{\rm i}\sdv_a}_{a=1}^{n_{2s}},\qquad
 \sdv_a=O(M^{-\infty}),&
 \label{DL_2s}
 \\
& \text{quartets:}\quad && \set{\clp_a^++{\rm i}\sdv_a,\clp*_a^++{\rm i}\sdv_a,\clp_a^--{\rm i}\sdv_a,\clp*_a^--{\rm i}\sdv_a}_{a=1}^{n_q},\qquad
\sdv_a=O(M^{-\infty}),&
 \label{DL_quart}
 \\
& \text{wide-pairs:}\quad&&
 \set{\wdp^+_a, \wdp*^-_a}_{a=1}^{n_w},&
 \label{DL_wp}
\end{alignat}
where $\sdv_a$ is called string deviation. We can combine all identifiers of non-real complex roots found in equations~\eqref{DL_2s}--\eqref{DL_wp} together to form a set of $\ho{n}=n_{2s}+2n_q+2n_w$ higher-level parameters
\begin{align*}
 \ho{\mu}=
 \set{\clp_a}_{a\leq n_{2s}} \cup
 \set{\clp_a,\clp*_a}_{a\leq n_q}
 \cup \set{\wdp_a,\wdp*_a}_{a\leq n_w}.
\end{align*}
We can show that parameters $\ho{\mu}$ are determined in terms of hole parameters $\set{\hle_a}$ through a set of equations \cite{DesL82} called higher level Bethe equations
\begin{gather}
 \aux*(\ho\mu_a)+1=0, \quad\forall a\leq \ho n,
\qquad
 \aux*(\la)=
 \prod_{a=1}^{n_h}\frac{\la-\hle_a-\frac{\rm i}{2}}{\la-\hle_a+\frac{\rm i}{2}}
 \prod_{a=1}^{\ho n}\frac{\la-\ho\mu_a+{\rm i}}{\la-\ho\mu_a-{\rm i}},
 \label{hlbe}
\end{gather}
while the density function for real roots is modified in the presence of complex roots as follows
\begin{gather*}
 \den_e(\la)=\den_g(\la)+\frac{1}{M}\sum_{a=1}^{n_h}\den_h(\la-\hle_a)+\frac{1}{M}\sum_{a=1}^{\ho n}\den*(\la-\ho\mu_a),\qquad
 \den*(\la)=-K_{\frac{1}{2}}(\la).
\end{gather*}
Equations \eqref{hlbe} are similar to the Bethe equations \eqref{Bethe_eq}, however they are inhomogeneous in nature. The rapidities of holes are inserted in these equations as inhomogeneity terms and they completely determine the positions of complex parameters.
The emergence of the higher level Bethe equations \eqref{hlbe} to describe complex roots is one of the most important finding of \cite{DesL82}, and of \cite{BabVV83} in a more general context of the $XXZ$ model.
This paper demonstrate that a similar emergent structure can also be found for the form factors of the excitations of Destri--Lowenstein type.
\par
Equations \eqref{hlbe} also determine the total spin or the eigenvalue of $S^z$ in terms of the cardinalities $n_h$, $\ho n$ of the holes and higher-level roots.
It can be shown that
\begin{align*}
s= \frac{1}{2}n_h-\ho n.
\end{align*}
With a fixed set of hole parameters $\set{\hle_a}_{a\leq n_h}$, one can further show that it leads to a $2^{n_h}$ dimensional degenerate subspace since equations \eqref{ene_eigen} and \eqref{mom_eigen} for the eigenvalues are also valid for the excitations containing complex roots.

\subsection{Form factors for the finite chain}
In this section we briefly review the determinant representations for the form factors that are obtained in the algebraic Bethe ansatz framework.
The main object that we want to consider is the longitudinal form factor
\begin{gather}
 |\mathcal{F}_z|^2=
\frac{\bra{\Psi_e}\sigma^z_m\ket{\Psi_g}\bra{\Psi_g}\sigma^z_m\ket{\Psi_e}} {\braket{\Psi_e}{\Psi_e}\braket{\Psi_g}{\Psi_g}}.
\label{ff_long_finite}
\end{gather}
In order to compute the matrix elements in this expression we should have the representation of local spin operators in the algebra of monodromy matrix elements.
This problem, known as quantum inverse scattering problem, was solved for spin 1/2 chains in \cite{KitMT99}, where the following relations were obtained
\begin{gather*}
 \sigma^z_m=
 \mathcal{T}^{m-1}\bigg(\frac{\rm i}{2}\bigg)\bigg\lbrace
 \mathcal{A}\bigg(\frac{\rm i}{2}\bigg)
 - \mathcal{D}\bigg(\frac{\rm i}{2}\bigg)
 \bigg\rbrace\mathcal{T}^{-m}\bigg(\frac{\rm i}{2}\bigg),
 \\[1ex]
 \sigma^+_m=
 \mathcal{T}^{m-1}\bigg(\frac{\rm i}{2}\bigg)
 \mathcal{C}\bigg(\frac{\rm i}{2}\bigg)
 \mathcal{T}^{-m}\bigg(\frac{\rm i}{2}\bigg),
 \\[1ex]
 \sigma^-_m=
 \mathcal{T}^{m-1}\bigg(\frac{\rm i}{2}\bigg)
 \mathcal{B}\bigg(\frac{\rm i}{2}\bigg)
 \mathcal{T}^{-m}\bigg(\frac{\rm i}{2}\bigg).
\end{gather*}
Using these relations, equation~\eqref{ff_long_finite} can be reduced to the ratio of scalar products.
A determinant representation for this type of scalar products was obtained by Slavnov \cite{Sla89}.
Let us summarise this result: let the set $\set{\la_1,\ldots,\la_N}$ be a solution of Bethe equations and $\set{\mu_1,\ldots,\mu_{N'}}$ be an arbitrary generic set of spectral parameters. Then the scalar product of corresponding states is given by
\begin{gather*}
 \braket{\Psi(\set{\la_a}_{a\leq N})}{\Psi(\set{\mu_a}_{a\leq N'})}
 = \delta_{N,N'} \frac{\prod_{j,k}(\mu_j-\la_k-{\rm i})}{\prod_{j<k}(\la_j-\la_k)(\mu_k-\mu_j)} \det\Sla,
 \\
 \Sla_{jk} = \aux(\mu_k)t(\mu_k-\la_j)-t(\la_j-\mu_k).
\end{gather*}
The norm of the Bethe states is given by the Gaudin determinant formula \cite{Gau83L,GauMcCW81,Kor82}
\begin{gather*}
 \braket{\Psi(\set{\la_a}_{a\leq N})}{\Psi(\set{\la_a}_{a\leq N})}
 = \frac{\prod_{j,k}(\la_j-\la_k-{\rm i})}{\prod_{j\neq k}(\la_j-\la_k)} \det\Gau,
 \\
 \Gau_{j,k}= \aux'(\la_j)\delta_{j,k}-2\pi {\rm i} K(\la_j-\la_k).
\end{gather*}
Before employing these determinant representations in equation~\eqref{ff_long_finite}, we must first recall that the $XXX$ model possesses an additional $\mathfrak{su}_2$ symmetry. In the scheme of multiplet decomposition~\eqref{mult_xxx} resulting from this symmetry, we can show that the only non-trivial value for the longitudinal form factors comes from the triplet excitations.
Moreover, the additional $\mathfrak{su}_2$ symmetry can be used to prove the following identities
\begin{gather}
 \bra{\Psi_1}\sigma^z_m\ket{\Psi_g}= -2\bra{\Psi_0}\sigma^+_m\ket{\Psi_g},
 \label{transmap_num1}
 \\
 \bra{\Psi_g}\sigma^z_m\ket{\Psi_1}= \bra{\Psi_g}\sigma^+_m\ket{\Psi_2},
 \label{transmap_num2}
 \\
 \braket{\Psi_1}{\Psi_1}=2\braket{\Psi_0}{\Psi_0}.
 \label{transmap_den}
\end{gather}
We can use these identities to rewrite the form factor \eqref{ff_long_finite} in terms of the matrix elements of off-diagonal local operators.
This technique however necessarily requires the use of Foda--Wheeler version of the Slavnov formula \cite{FodW12}
\begin{gather*}
 \braket{\Psi_\ell(\set{\la}_{N})}{\Psi(\set{\mu}_{N+\ell})}=
 \frac{(-1)^{N\ell+\frac{\ell^2}{2}}\ell!\prod_{k=1}^{N+\ell}q(\mu_k-{\rm i})}
 {\prod_{j<k}^{N}(\la_j-\la_k)\prod_{j>k}^{N+\ell}(\mu_k-\mu_j)} \det\Sla^{(\ell)},
 \\
 \Sla^{(\ell)}_{j,k}= (\Sla \mid \FW)^T,
 \\
 \FW_{a,k}= \aux(\mu_k)(\mu_k+{\rm i})^{a-1}-\mu_k^{a-1}, \qquad
 a=1,\ldots,\ell.
\end{gather*}
Therefore, we have seen that the form factors \eqref{ff_long_finite} can be expressed as the ratio of determinants that have infinite size in the thermodynamic limit with a rational prefactor
\begin{gather}
 |F_{z}|^2 = -2
\pl_{j=1}^{\frac{M}{2}-1}\frac{q_g(\mu_j-{\rm i})}{q_e(\mu_j-{\rm i})}\pl_{k=1}^{ \frac{M}{2}}\frac{q_e(\la_k-{\rm i})}{q_g(\la_k-{\rm i})}
\frac{\det\Sla_g}{\det\Gau_g} \frac{\det\big(\Sla_e\big|\FW\big)}{\det\Gau_e} ,
 \label{det_rep_finite}
\end{gather}
where $q_g$ and $q_e$ are Baxter polynomials \eqref{baxq} for the ground state and the excited state.

Our goal is to compute the form factors in the thermodynamic limit where the matrices in this determinant representation become infinite dimensional.
In this limit, the close-pair roots also condense to form 2-string or quartet configuration.
As a result, the auxiliary function (\ref{exp_count}) develops a singularity as it contains a pole in the string deviation parameter.

Since we are using the Destri--Lowenstein picture (see equations~\eqref{DL_rl}--\eqref{DL_wp}) for excitations, we will separately look at columns of the both Slavnov matrices classified into real, close-pair and wide-pair blocks.
First $n_r+1$ columns of Slavnov matrix $\Sla_g$ containing `real' roots can be written as
\begin{gather}
 \Sla^{\rlx}_{g;jk}=
 \aux_g(\check\rl_k)t(\check\rl_k-\la_j)-t(\la_j-\check\rl_k).
 \label{sla_mat1_r}
\end{gather}
Note that we have enlarged the set of real roots to include a special parameter $\frac{\rm i}{2}$ by imposing the following notation
\begin{gather*}
 \set{\check\rl_j}_{j\leq n_r+1} =
 \set{\rl_j}_{j\leq n_r}\cup\bigg\{\frac{\rm i}{2}\bigg\}.
\end{gather*}

When it comes to the next block of columns containing close-pair roots
\begin{gather*}
 \Sla^{\clpx \pm}_{g;ja}=
 \aux_g(\clp^\pm_a\pm {\rm i}\sdv_a)t(\clp^\pm_a-\la_j\pm {\rm i}\sdv_a)-t(\la_j-\clp^\pm_a\mp {\rm i}\sdv_a)
\end{gather*}
we find that half of these columns have a pole in the string deviation parameter $\sdv_a$.
However, it can be properly regularised at the level of determinant using the following two relations
\begin{gather*}
 \aux_g(\clp^+_a+{\rm i}\sdv_a)\aux_g(\clp^-_a-{\rm i}\sdv_a)=1+O(\sdv_a),
\\
 t(\clp^-_a-\la)=t(\la-\clp^+_a)
\end{gather*}
to recombine the columns as follows
\begin{gather*}
 \Sla^{\clpx-}_{g;j,a}\leftarrow
 \Sla^{\clpx-}_{g;j,a}+\aux_g(\clp^-_a-{\rm i}\sdv_a)\Sla^{\clpx+}_{g;j,a}.
\end{gather*}
This gives us a regularised expressions for the columns $\Sla^{\clpx-}_g$
\begin{align}
\Sla^{\clpx-}_{g;j,a}={}&\aux_{g}(\clp^-_a-{\rm i}\sdv_a)
\lbrace t(\clp^-_a-\la_j-{\rm i}\sdv_a)-t(\la_j-\clp^+_a-{\rm i}\sdv_a)\rbrace \nonumber
 \\
&+2\pi {\rm i}\lbrace K(\la_j-\clp^+_a)-K(\la_j-\clp^-_a)\rbrace +O(\sdv_a),
 \label{sla_mat1_clp-}
\end{align}
while the remaining close-pair columns can be truncated to the leading order in $\sdv_a$
\begin{align}
 \Sla^{\clpx +}_{g;j,a}=
 -t(\la_j-\clp^+_a)+O(\sdv_a).
 \label{sla_mat1_clp+}
\end{align}

Since the exponential counting function is constant in the region where the wide-pairs are located
\begin{gather*}
 \aux_g(z)=1,\qquad |\operatorname{Im} z|>1,
\end{gather*}
here we can write the columns of $\Sla_g$ containing wide-pairs as simply
\begin{gather}
\begin{pmatrix}
\Sla^{\wdpx +}_{g;ja} \\[\jot] \Sla^{\wdpx -}_{g;ja}
\end{pmatrix}
= \begin{pmatrix}
 t(\wdp^+-\la_j) \\ t(\wdp*^--\la_j)
 \end{pmatrix}
 - \begin{pmatrix}
 t(\la_j-\wdp^+) \\ t(\la_j-\wdp*^-)
 \end{pmatrix}\!.
 \label{sla_mat1_wdp}
\end{gather}
Therefore we say that the total matrix $\Sla_g$ has block decomposition given by equations~\eqref{sla_mat1_r}, \eqref{sla_mat1_clp+}, \eqref{sla_mat1_clp-}, and \eqref{sla_mat1_wdp} arranged into the columns
\begin{gather*}
\Sla_g= \big(\Sla_g^{\rlx}\mid \Sla_g^{\clpx+}
\mid \Sla_g^{\clpx-}\mid \Sla_g^{\wdpx+}\mid \Sla_g^{\wdpx-}\big).
\end{gather*}

Let us now turn to the rectangular Slavnov matrix $\Sla_e$ in equation~\eqref{det_rep_finite}.
It has columns that can be organised into blocks corresponding to real, close-pair and wide-pair roots
\begin{gather*}
 \Sla_{e;j,k}^{\rlx}=
 \aux_e\big(\check\la_j\big)t\big(\check\la_j-\rl_k\big)-t\big(\rl_k-\check\la_j\big),
 \\
 \Sla_{e;j,a}^{\clpx+}=
 \aux_e\big(\check\la_j\big)t\big(\check\la_j-\clp^+_a\big)-t\big(\clp^+_a-\check\la_j\big),
 \\
 \Sla_{e;j,a}^{\clpx-}=
 \aux_e\big(\check\la_j\big)t\big(\check\la_j-\clp^-_a\big)-t\big(\clp^-_a-\check\la_j\big),
 \\
 \Sla_{e;j,a}^{\wdpx+}=
 \aux_e\big(\check\la_j\big)t\big(\check\la_j-\wdp^+_a\big)-t\big(\wdp^+_a-\check\la_j\big),
 \\
 \Sla_{e;j,a}^{\wdpx-}=
 \aux_e\big(\check\la_j\big)t\big(\check\la_j-\wdp*^-_a\big)-t\big(\wdp*^-_a-\check\la_j\big).
\end{gather*}
In this case, the extra parameter gets added to the set of ground state roots since we have taken the right action of the operator $\sigma^+_m$ in equation~\eqref{transmap_num2}.
\begin{gather*}
 \big\{\check\la_j\big\}_{j\leq \frac{M}{2}+1}=\set{\la_j}_{j\leq \frac{M}{2}}\cup\bigg\{\frac{\rm i}{2}\bigg\}.
\end{gather*}
Note that unlike the Slavnov matrix $\Sla_g$, there are no singular terms arising from close-pairs in the matrix $\Sla_e$.
However we do encounter a similar issue at the level of Gaudin matrix for the norm and the prefactor to its determinant. As we do not use the string hypothesis we will not use here the form of the Gaudin determinant with complex roots obtained in \cite{KirK88}.
Due to the nature of this term, we judge it suitable to regularise it later.

\section{Gaudin extraction}
\label{sec:gauex}
We use our method of the \emph{Gaudin matrix extraction} that was originally introduced in \cite{KitK19} for the factorisation problem in equation~\eqref{det_rep_finite}.

\begin{subequations}
The matrices resulting from this extraction are given by actions of inverse Gaudin matrices on the respective Slavnov matrices for the each fraction of determinants
\begin{gather}
 \Fmat_{g}= \Gau_g^{-1} \Sla_g,
 \label{gauex_mat1}
\\
 \big(\Fmat_{e}\mid\FW\big)^T=
 \begin{pmatrix}
 \Gau_{\rm e}^{-1} & \\ & \Id_2
 \end{pmatrix}
 \big(\Sla_{e}\mid\FW\big)^{T}.
 \label{gauex_mat2}
\end{gather}
\end{subequations}
We note that Foda--Wheeler block remains invariant under this procedure of Gaudin extraction and hence it will not enter our discussion in this paper.
The matrices $\Fmat_g$ and $\Fmat_e$ solves the system of linear equations respectively
\begin{subequations}
\label{linsys}
\begin{gather}
 \aux'_g(\la_j)\Fmat_{g;j,k}
 -2\pi {\rm i}\sum_{l=1}^{\frac{M}{2}}K(\la_j-\la_l)\Fmat_{g;l,k}
=\aux_g(\check\mu_k)t(\check\mu_k-\la_j)-t(\la_j-\check\mu_k),
 \label{linsys_mat1}
 \\
 \aux'_e(\mu_k)\Fmat_{e;j,k}
 -2\pi {\rm i}\sum_{l=1}^{\frac{M}{2}-1}K(\mu_k-\mu_l)\Fmat_{e;k,l}
=\aux_e\big(\check\la_j\big)\big(\check\la_j-\mu_k\big)-t\big(\mu_k-\check\la_j\big).
 \label{linsys_mat2}
\end{gather}
\end{subequations}
In this section we will compute the resulting infinite matrices $\Fmat_g$ and $\Fmat_e$ by solving integral equations in thermodynamic limit.
As this procedure is slightly different for two matrices, we will treat them separately, starting with the matrix $\Fmat_g$.
In both cases, a special attention needs to be paid to the effect of complex roots that are made up of close-pairs and wide-pairs in the Destri--Lowenstein picture.

\subsection[Integral equations for the matrix \protect{$F\textunderscore{}g$}]
{Integral equations for the matrix $\boldsymbol{{\cal F}_g}$}

We decompose the matrix $\Fmat_g$ into the following blocks
\begin{gather*}
 \Fmat^r_{g;j,k}=\Fmat_{g;j,k},\qquad k\leq n_r+1,
 \\
 \Fmat^{\clpx +}_{g;j,a}=\Fmat_{g;j,n_\rlx+a+1},\qquad a\leq n_{\clpx},
 \\
 \Fmat^{\clpx-}_{g;j,a}=\Fmat_{g;j,n_\rlx+n_\clpx+a+1},\qquad a\leq n_{\clpx},
 \\
 \Fmat^{\wdpx +}_{g;j,a}=\Fmat_{g;j,n_\rlx+2n_\clpx+a+1},\qquad a\leq n_{\wdpx},
 \\
 \Fmat^{\wdpx-}_{g;j,a}=\Fmat_{g;j,n_\rlx+2n_\clpx+n_\wdpx+a+1},\qquad a\leq n_{\wdpx}.
\end{gather*}
The system of linear equations for the block $\Fmat^r_g$ obtained from equation~\eqref{gauex_mat1} reads
\begin{gather}
 \aux_g'(\la_j)\Fmat^{r}_{g;j,k}
 -2\pi {\rm i}\sum_{l=1}^{\frac{M}{2}}K(\la_j-\la_l)\Fmat^{r}_{g;l,k}
 = \aux_g(\rl_k)t(\rl_k-\la_j)-t(\la_j-\rl_k).
 \label{linsys_mat1_rl}
\end{gather}
Here we will follow exactly the same procedure as in \cite{KitK19}, we write down an auxiliary meromorphic function $G_g\colon \Cset^2\to\Cset$ with the property
\begin{gather*}
 G_{g}(\la_j,\rl_k)= \aux'_g(\rl_j)\Fmat_{g;j,k}.
\end{gather*}
We replace the sum in equation~\eqref{linsys_mat1_rl} as integral over the contour going around the real line. Since the auxiliary function $\aux_g$ can be estimated in the thermodynamic to be exponentially small (large) over (below) the real line, we can write the following integral equation for the func\-tion~$G_g$
\begin{gather}
G_{g}(\la_j,\check\rl_k)
+\int_{\Rset+{\rm i}\alpha} K(\la_j-\tau) G_g(\tau,\check\rl_k) {\rm d}\tau
= (1+\aux_g(\check\rl_k))t(\check\rl_k-\la_j).
\label{intsys1_rl}
\end{gather}
For more clarification, this generalised condensation property is discussed in Appendix~\ref{app:gen_condn}.
Note that this has been extended to include the extra parameter $\check\rl_{n_r+1}=\frac{\rm i}{2}$.
Since $\aux_g\big(\frac{\rm i}{2}\big)=0$ we recover an integral equation identical to equation~\eqref{intsys1_rl} obtained for all real roots.
The solution of this integral equation returns the Cauchy elements of the matrix $\Fmat_g$ which are given by
\begin{gather}
\Fmat_{g;j,k}^{r}=
\frac{1+\aux_g(\check\rl_k)}{\aux_g'(\la_j)}
\frac{\pi}{\sinh\pi(\check\rl_k-\la_j)}.
\label{inteq1_rl_sol}
\end{gather}
Until this point the computation is identical to the two-spinon example \cite{KitK19}. We obtain a Cauchy matrix from this procedure, however in this example it turns out to be a rectangular matrix.
Remaining matrix blocks which arise due to the presence of complex roots $\Fmat^{\clpx\pm}$ and $\Fmat^{\wdpx\pm}$ are the primary point of investigation of this paper. We find that its elements deviate from the Cauchy matrix.

We obtain the following system of equations for blocks $\Fmat^{\clpx\pm}$ coming from the close-pairs.
\begin{gather}
 \aux_g'(\la_j)\Fmat^{\clpx+}_{g;j,a}
 -2\pi {\rm i}\sum_{l=1}^{\frac M2}K(\la_j-\la_l)\Fmat^{\clpx+}_{g;j,l}
 = -t(\la_j-\clp^+_a),
 \label{linsys1_clp+}
\\
 \aux_g'(\la_j)\Fmat^{\clpx-}_{g;j,a}
 \!-2\pi {\rm i}\!\sum_{l=1}^{\frac M2}\!K(\la_j\!-\la_l)\Fmat^{\clpx-}_{g;j,l} =
 \aux_g(\clp^-_a\!-{\rm i}\sdv_a)\lbrace
 t(\clp^-_a\!-\la_j\!-{\rm i}\sdv_a)\!-t(\la_j\!-\clp^+_a\!-{\rm i}\sdv_a) \rbrace \nonumber
 \\ \hphantom{ \aux_g'(\la_j)\Fmat^{\clpx-}_{g;j,a}
 \!-2\pi {\rm i}\sum_{l=1}^{\frac M2}K(\la_j\!-\la_l)\Fmat^{\clpx-}_{g;j,l} =}
 {}-2\pi {\rm i}\lbrace K(\la_j-\clp^+_a)-K(\la_j-\clp^-_a) \rbrace.
 \label{linsys1_clp-}
\end{gather}
The regular condensation property can be used to convert the sum in these equations as integrals.
Thus from equation~\eqref{linsys1_clp+} we get the following integral equation for the function $G_g$
\begin{gather*}
 G_g(\la,\clp^+_a)+\int_{\Rset}K(\la-\tau)G_g(\tau,\clp^+_a)\,{\rm d}\tau = -t(\la-\clp^+_a),
\end{gather*}
admitting the solution
\begin{gather}
 G_g(\la,\clp^+_a)=\den_{\frac12}(\la,\clp_a).
 \label{inteq1_clp+_sol}
\end{gather}
On the other hand we get the following integral equation from equation~\eqref{linsys1_clp-}
\begin{gather*}
 G_g(\la,\clp^-_a)+\int_{\Rset}K(\la-\tau)G_g(\tau-\clp^-_a){\rm d}\tau
 \\ \qquad
{} =
 a_g(\clp^-_a-{\rm i}\sdv_a)\lbrace t(\clp^-_a-\la-\sdv_a)-t(\la-\clp^+_a-{\rm i}\sdv_a)\rbrace
 -2\pi {\rm i}\lbrace K(\la-\clp^+_a)-K(\la-\clp^-_a) \rbrace .
\end{gather*}
Solving this integral equation gives us only a partial Cauchy matrix
\begin{align}
 G_g(\la,\clp^-_a)={}&
 \aux_g(\clp^-_a-{\rm i}\sdv_a)\big\lbrace
 \den_{\frac12}(\la,\clp_a+{\rm i}\sdv_a)-\den_{\frac12}(\la_j,\clp_a-{\rm i}\sdv_a) \big\rbrace\nonumber
 \\
 &+2\pi {\rm i}\lbrace \den_1(\la,\clp^+_a)-\den_1(\la,\clp^-_a)\rbrace.
 \label{inteq1_clp-_sol}
\end{align}
Similarly, integral equations for the wide-pairs read
\begin{gather*}
 \begin{pmatrix} G_g(\la,\wdp^+_a) \\ G_g(\la,\wdp*^-_a) \end{pmatrix}
 +\int_{\Rset}K(\la-\tau)
 \begin{pmatrix} G_g(\tau,\wdp^+_a) \\ G_g(\tau,\wdp*^-_a) \end{pmatrix}
 {\rm d}\tau =
 \begin{pmatrix} t(\wdp^+_a-\la_j) \\ t(\wdp*^-_a-\la_j) \end{pmatrix}
 - \begin{pmatrix} t(\la_j-\wdp^+_a) \\ t(\la_j-\wdp*^-_a) \end{pmatrix}
\end{gather*}
admitting solutions
\begin{gather}
 \begin{pmatrix} G_g(\la,\wdp^+_a) \\ G_g(\la,\wdp*^-_a) \end{pmatrix}
 = 2\pi {\rm i}
 \begin{pmatrix}
 \den_{\frac12}(\la,\wdp_a+{\rm i})-\den_{\frac12}(\la,\wdp_a)
 \\
 \den_{\frac12}(\la,\wdp*_a)-\den_{\frac12}(\la,\wdp*_a-{\rm i})
 \end{pmatrix}\! .
 \label{inteq1_wdp_sol}
\end{gather}
From equations~\eqref{inteq1_rl_sol}, \eqref{inteq1_clp+_sol}, \eqref{inteq1_clp-_sol} and \eqref{inteq1_wdp_sol} we obtain the components of the matrix $\Fmat_g$
\begin{gather*}
 \Fmat_{g;j,k}^{\rcx}= \frac{1+\aux(\check\rl_k)}{\aux_g'(\la_j)}
 \frac{\pi}{\sinh\pi\big(\check\rl^+_k-\la_j\big)},
 \\
 \Fmat^{c-}_{g;j,a}=
 \frac{\aux_g(\clp^-_a-{\rm i}\sdv_a)}{\aux_g'(\la_j)}
\bigg\lbrace \frac{\pi}{\sinh\pi(\clp^+_a-\la_j+{\rm i}\sdv_a)}
 + \frac{\pi}{\sinh\pi(\clp^-_a-\la_j-{\rm i}\sdv_a)} \bigg\rbrace
 \\ \hphantom{ \Fmat^{c-}_{g;j,a}=}
 {}+\frac{2\pi {\rm i}}{\aux_g'(\la_j)}
 \big\lbrace \den_{1}(\la_j,\clp^+_a)-\den_{1}(\la_j,\clp^-_a) \big\rbrace,
 \\
 \Fmat_{g;j,a}^{\wdpx+}= \frac{2\pi {\rm i}}{\aux_g'(\la_j)}
 \big\lbrace \den_{\frac12}(\la_j,\wdp_a+{\rm i})-\den_{\frac12}(\la_j,\wdp_a) \big\rbrace,
 \\
 \Fmat_{g;j,a}^{\wdpx-}= \frac{2\pi {\rm i}}{\aux_g'(\la_j)}
 \big\lbrace \den_{\frac12}(\la_j,\wdp*_a)-\den_{\frac12}(\la_j,\wdp*_a-{\rm i}) \big\rbrace,
\end{gather*}
where in $\Fmat_{g}^{\rcx}$ we have combined all the Cauchy terms together spanning over the set obtained by the union
\begin{gather*}
 \set{\check\rl^+_a}_{a\leq n_r+n_\clpx+1}
 = \set{\rl_a}_{a\leq n_r}\cup \bigg\{\frac{\rm i}{2}\bigg\}\cup \set{\clp^+_a}_{a\leq n_\clpx}.
\end{gather*}
This follows from the fact that half of the close-pair columns behave similarly to the real columns. This breaking apart of close-pairs is an important observations
 and it also holds true for $\Fmat_e$.

\subsection[Integral equations for the matrix \protect{$F\textunderscore{}e$}]
{Integral equations for the matrix $\boldsymbol{{\mathcal F}_e}$}

We apply the decomposition of the matrix $\Fmat_e$ into the blocks of columns by following the Destri--Lowenstein picture
\begin{gather}
\Fmat^{\rlx}_{e;k,j}=\Fmat_{e;k,j}, \qquad j\leq n_{\rlx},
\label{mat2_rl}
\\
\Fmat^{\clpx +}_{e;k,a}=\Fmat_{e;k,n_{\rlx}+a},\qquad a\leq n_{\clpx},
\label{mat2_cp+}
\\
\Fmat^{\clpx -}_{e;k,a}=\Fmat_{e;k,n_{\rlx}+n_{\clpx}+a},\qquad a\leq n_{\clpx},
\label{mat2_cp-}
\\
\Fmat^{\wdpx +}_{e;k,a}=\Fmat_{e;k,n_{\rlx}+2n_{\clpx}+a},\qquad a\leq n_{\wdpx},
\label{mat2_wp+}
\\
\Fmat^{\wdpx -}_{e;k,a}=\Fmat_{e;k,n_{\rlx}+2n_{\clpx}+n_{\wdpx}+a}, \qquad a\leq n_{\wdpx}.
\label{mat2_wp-}
\end{gather}
The system of equation \eqref{linsys_mat2} for the block of real columns $\Fmat^{\rlx}_e$ \eqref{mat2_rl} can be rewritten as
\begin{gather}
\aux'_e(\rl_j)\Fmat^{\rlx}_{e;k,j} -
 2\pi {\rm i}\sum_{l=1}^{n_r} K(\rl_j-\rl_l) \Fmat^{\rlx}_{e;k,l}\nonumber
 \\ \qquad
 {}= \aux_e\big(\check\la_k\big)t\big(\check\la_k-\rl_j\big)-t\big(\rl_j-\check\la_k\big)
+2\pi {\rm i}\sum_{a=1}^{n_\clpx}
 \big\lbrace K(\rl_j-\clp^+_a)\Fmat^{\clpx +}_{e;k,a} + K(\rl_j-\clp^-_a)\Fmat^{\clpx-}_{e;k,a}\big\rbrace\nonumber
 \\ \qquad\phantom{=}
{}+2\pi {\rm i}\sum_{a=1}^{n_\wdpx}
 \big\lbrace K(\rl_j-\wdp^+_a)\Fmat^{\wdpx +}_{e;k,a} + K(\rl_j-\wdp*^-_a)\Fmat^{\wdpx-}_{e;k,a}\big\rbrace.
 \label{linsys2_rl}
\end{gather}
Let us follow exactly the same procedure as before, we write down an auxiliary meromorphic function $G_e\colon \Cset^2\to\Cset$ with the following property
\begin{gather*}
 G_{e}\big(\rl_j,\check\la_k\big)= \aux'_e(\rl_j)\Fmat_{e;k,j}.
\end{gather*}
A generalised condensation property (Appendix \ref{app:gen_condn}) can be applied here again. Let us first convert the sum on the left hand side of equation~\eqref{linsys2_rl} into a contour integral
\begin{gather*}
 2\pi {\rm i}\sum_{l=1}^{\frac{M}{2}-1} K(\rl_j-\rl_l)\Fmat^{\rlx}_{e;k,l}
 = \bigg(\int_{\Rset-{\rm i}\alpha}-\int_{\Rset+{\rm i}\alpha} \bigg)
 K(\rl_j-\nu)\frac{G_e\big(\nu,\check\la_k\big)}{1+\aux_e(\nu)}{\rm d}\nu,
\end{gather*}
where the integrals on edges of the contour at infinity vanishes as the integrand decays for large $\nu$ for any function $G_e$ that is bounded at infinity.
Furthermore, we can use the fact that exponential counting function gets exponentially small and exponentially large in $M$ in the corresponding half-planes,
to obtain integral equation for the auxiliary function $G_e$\vspace{-1ex}
\begin{gather}
 G_e\big(\rl,\check\la_k\big)
 +\int_{\Rset+{\rm i}\alpha}K(\rl-\nu)G_e\big(\nu,\check\la_k\big){\rm d}\nu\nonumber
 \\[-.5ex] \qquad
 {}= \big(1+\aux_e\big(\check\la_k\big)\big)t\big(\check\la_k-\rl\big)
 -2\pi {\rm i}\sum_{a=1}^{n_h}
 K(\rl-\hle_a)\frac{G_e\big(\hle_a,\check\la_k\big)}{\aux'_e(\hle_a)}\nonumber
\\[-.5ex] \qquad\hphantom{=}
{}+2\pi {\rm i}\sum_{a=1}^{n_\clpx}
 \big\lbrace K(\rl_j-\clp^+_a)\Fmat^{\clpx +}_{e;k,a} + K(\rl_j-\clp^-_a)\Fmat^{\clpx -}_{e;k,a}\big\rbrace\nonumber
 \\[-.5ex] \qquad\hphantom{=}
{}+2\pi {\rm i}\sum_{a=1}^{n_\wdpx}
\big\lbrace K(\rl_j-\wdp^+_a)\Fmat^{\wdpx +}_{e;k,a} + K(\rl_j-\wdp*^-_a)\Fmat^{\wdpx -}_{e;k,a}\big\rbrace .
 \label{intsys2_rl}
\end{gather}
Let us now turn our attention to blocks $\Fmat^{\clpx\pm}_e$ of columns for close-pairs. In this case, we can write the system equation~\eqref{linsys_mat2} altogether as\vspace{-1ex}
\begin{gather}
 \begin{pmatrix}
 \aux'_e(\clp^+_a)\Fmat^{\clpx +}_{e;k,a}
 -2\pi {\rm i} K(2{\rm i}\sdv_a)\Fmat^{\clpx -}_{e;k,a}
 \\
 \aux'_e(\clp^-_a)\Fmat^{\clpx -}_{e;k,a}
 -2\pi {\rm i} K(-2{\rm i}\sdv_a)\Fmat^{\clpx +}_{e;k,a}
 \end{pmatrix}
 -2\pi {\rm i}\sum_{l=1}^{n_\rlx}
 \begin{pmatrix}
 K(\clp^+_a-\rl_l+{\rm i}\sdv_a)\Fmat_{e;k,l} \\ K(\clp^-_a-\rl_l-{\rm i}\sdv_a)\Fmat_{e;k,l}
 \end{pmatrix}\nonumber
\\ \quad
 {}-2\pi {\rm i} \sum_{\underset{b\neq a}{b=1}}^{n_\clpx}
 \begin{pmatrix}
 K(\clp_a-\clp_b+{\rm i}+{\rm i}(\sdv_a+\sdv_b))\Fmat^{\clpx -}_{e;k,b}
 \\
 K(\clp_a-\clp_b-{\rm i}-{\rm i}(\sdv_a+\sdv_b))\Fmat^{\clpx +}_{e;k,b}
 \end{pmatrix}
 -2\pi {\rm i} \sum_{b=1}^{n_{\clpx}}
 \begin{pmatrix}
 K(\clp_a-\clp_b+{\rm i}(\sdv_a-\sdv_b))\Fmat^{\clpx +}_{e;k,b}
 \\
 K(\clp_a-\clp_b-{\rm i}(\sdv_a-\sdv_b))\Fmat^{\clpx -}_{e;k,b}
 \end{pmatrix}\nonumber
\\ \quad
 {}-2\pi {\rm i}\sum_{b=1}^{n_{\wdpx}}
 \begin{pmatrix}
 K(\clp_a-\wdp_b+{\rm i}+{\rm i}\sdv_a)\Fmat^{\wdpx -}_{e;k,b}
 \\
 K(\clp_a-\wdp*_b-{\rm i}-{\rm i}\sdv_a)\Fmat^{\wdpx +}_{e;k,b}
 \end{pmatrix}
 -2\pi {\rm i}\sum_{b=1}^{n_{\wdpx}}
 \begin{pmatrix}
 K(\clp_a-\wdp_b+{\rm i}\sdv_a)\Fmat^{\wdpx +}_{e;k,b}
 \\
 K(\clp_a-\wdp*_b-{\rm i}\sdv_a)\Fmat^{\wdpx -}_{e;k,b}
 \end{pmatrix}\nonumber
\\
{}=\aux_e\big(\check\la_k\big)
\begin{pmatrix} t\big(\check\la_k-\clp^+_a\big) \\ t\big(\check\la_k-\clp^-_a\big) \end{pmatrix}
 -\begin{pmatrix} t\big(\clp^+_a-\check\la_k\big) \\ t\big(\clp^-_a-\check\la_k\big) \end{pmatrix}\! .
 \label{linsys2_clp}
\end{gather}
In the presence of close-pairs, the Gaudin matrix $\Gau_e$ contains singular terms. In particular they appear in the derivative of the counting function\vspace{-1ex}
\begin{gather*}
 \aux'_{e}(\clp^\pm_a)= \reg \aux'_e(\clp^\pm) +2\pi {\rm i}K(\pm 2{\rm i}\sdv_a),
\end{gather*}
where $K(\pm 2{\rm i}\sdv_a)$ contains simple poles in $\sdv_a$ with same sign for both residues since the function~$K$ is even.
Therefore the rearrangement of singular terms in equation~\eqref{linsys2_clp} allows us to see that there is a pairwise degeneracy in close-pair columns.\vspace{-1ex}
\begin{gather}
 \Fmat^{\clpx +}_{e;j,a}-\Fmat^{\clpx -}_{e;j,a}
 = (2{\rm i}\sdv_a)\Fmat^{\clpx\Delta}_{e;j,a}.
 \label{mat2_clp_diff}
\end{gather}
Let us therefore recombine close-pair columns of the matrix $\Fmat_e$, so that $\Fmat^{\clpx +}_{e}$ is replaced by the regularised difference $\Fmat^{\clpx\Delta}_{e}$ from equation~\eqref{mat2_clp_diff}, where the string deviation term thus extracted from the determinant of $\Fmat_e$ can be used to regularise the same simple divergence in $\sdv_a$ in its prefactor.

We can now write down the following equations for the recombined close-pair columns
\begin{align}
 \Fmat^{\clpx\Delta}_{j,a} ={}&
 \big(1+\aux_e\big(\check\la_j\big)\big)t\big(\check\la_j-\clp^+_a\big)
 -\!\!\int_{\Rset+{\rm i}\alpha}\!\!\!\!\!\! K(\clp^+_a\!-\nu)G_e\big(\nu,\check\la_j\big){\rm d}\nu
 \!- \sum_{b=1}^{n_h} K(\clp^+_a\!-\hle_b)
 \frac{2\pi {\rm i}G_e\big(\hle_b,\check\la_j\big)}{\aux'_e(\hle_b)}\nonumber
 \\[-.5ex]
 & -(\reg\aux'_e(\clp^+_a)-2\pi {\rm i}K(0))\Fmat^{\clpx -}_{e;j,a}
 +2\pi {\rm i} \sum_{b\neq a}^{n_{\clpx}}
 K^{(c)}\bigg(\clp_a-\clp_b+\frac{\rm i}{2}\bigg)\Fmat^{\clpx -}_{e;j,a}\nonumber
\\[-.5ex]
& +2\pi {\rm i}\sum_{b=1}^{n_{\wdpx}}
 K(\clp_a-\wdp_b)\Fmat^{\wdpx +}_{e;j,a}
 +2\pi {\rm i}\sum_{b=1}^{n_{\wdpx}}
 K(\clp_a-\wdp*_b+{\rm i})\Fmat^{\wdpx -}_{e;j,a}.
 \label{intsys2_cpdiff}
\end{align}
Meanwhile we can eliminate the difference of columns terms $\Fmat^{\clpx\Delta}_e$ in the pair of equations \eqref{linsys2_clp} by adding them together. This leads us to the system of equations
\begin{align}
\aux*'(\clp_a)\Fmat^{\clpx -}_{e;j,a} &-2\pi {\rm i}\sum_{b=1}^{n_{\clpx}}
\lbrace 2K(\clp_a-\clp_b)+K_{2}(\clp_a-\clp_b)\rbrace\Fmat^{\clpx -}_{e;j,b}\nonumber
\\
&-2\pi {\rm i}\sum_{b=1}^{n_\wdpx}K^{(c)}\bigg(\clp_a-\wdp_b-\frac{\rm i}{2}\bigg)\Fmat^{\wdpx +}_{e;j,b}
-2\pi {\rm i}\sum_{b=1}^{n_\wdpx}K^{(c)}\bigg(\clp_a-\wdp*_b+\frac{\rm i}{2}\bigg)\Fmat^{\wdpx -}_{e;j,b}\nonumber
\\
 &= 2\pi {\rm i} \big(1+\aux_e\big(\check\la_j\big)\big)K\bigg(\clp_a-\check\la_j+\frac{\rm i}{2}\bigg)
-\int_{\Rset+{\rm i}\alpha}K^{(c)}(\clp_a-\nu)G_e\big(\nu,\check\la_j\big){\rm d}\nu\nonumber
\\
& \phantom{=}-2\pi {\rm i}\sum_{b=1}^{n_h} {K^{(c)}(\clp_a-\hle_b)}
 \frac{G_e\big(\hle_b,\check\la_j\big)}{\aux'_e(\hle_b)}.
 \label{intsys2_cpsum}
\end{align}
Here we have used the short-hand notation $K^{(c)}$ in equations~\eqref{intsys2_cpsum} and~\eqref{intsys2_cpdiff} that stands for the combination
\begin{gather*}
 K^{(c)}(\la)=K\bigg(\la-\frac{\rm i}{2}\bigg)+K\bigg(\la+\frac{\rm i}{2}\bigg).
\end{gather*}
The derivative of the higher-level auxiliary function $\aux*'$ \eqref{hlbe} is obtained from the combination
\begin{gather*}
 \aux*'(\clp_a)= \reg\aux'(\clp^+_a)+\reg\aux'(\clp^-_a).
\end{gather*}
On the contrary for the wide-pairs we directly obtain the relations
\begin{gather*}
 \aux*'(\wdp_a)=\aux'(\wdp^+),\qquad
 \aux*'(\wdp*_a)=\aux'(\wdp*^-)
\end{gather*}
that allows us to write
\begin{align}
\aux*'(\wdp_a)\Fmat^{\wdpx +}_{e;j,a}
& -2\pi {\rm i}\sum_{b=1}^{n_{\clpx}}K^{(c)}\bigg(\wdp_a-\clp_b+\frac{\rm i}{2}\bigg)\Fmat^{\clpx -}_{e;j,b}\nonumber
\\
&-2\pi {\rm i}\sum_{b=1}^{n_{\wdpx}}K(\wdp_a-\wdp_b)\Fmat^{\wdpx +}_{e;j,a}
-2\pi {\rm i}\sum_{b=1}^{n_{\wdpx}}K(\wdp_a-\wdp*_b+{\rm i})\Fmat^{\wdpx -}_{e;j,a}\nonumber
\\
& =\big(1+\aux_e\big(\check\la_j\big)\big)t\bigg(\check\la_j-\wdp_a-\frac{\rm i}{2}\bigg)
-\int_{\Rset+{\rm i}\alpha} K\bigg(\wdp_a-\nu+\frac{\rm i}{2}\bigg)G_e\big(\nu,\check\la_j\big){\rm d}\nu\nonumber
\\
&\phantom{=}-2\pi {\rm i}\sum_{b=1}^{n_h} {K\bigg(\wdp_a-\hle_b+\frac{\rm i}{2}\bigg)}
\frac{G_e\big(\hle_b,\check\la_j\big)}{\aux'_e(\hle_b)}
\label{intsys2_wp+}
\end{align}
and
\begin{align}
\aux*'(\wdp*_a)\Fmat^{\wdpx -}_{e;j,a} &-2\pi {\rm i}\sum_{b=1}^{n_{\clpx}}K^{(c)} \bigg(\wdp*_a-\clp_b+\frac{\rm i}{2}\bigg)\Fmat^{\clpx -}_{e;j,b}\nonumber
\\
&-2\pi {\rm i}\sum_{b=1}^{n_{\wdpx}}K(\wdp*_a-\wdp_b-{\rm i})\Fmat^{\wdpx +}_{e;j,a}
-2\pi {\rm i}\sum_{b=1}^{n_{\wdpx}}K(\wdp*_a-\wdp*_b)\Fmat^{\wdpx -}_{e;j,a}\nonumber
\\
&=\big(1+\aux_e\big(\check\la_j\big)\big)t\bigg(\check\la_j-\wdp*_a+\frac{\rm i}{2}\bigg)
-\int_{\Rset+{\rm i}\alpha}K\bigg(\wdp*_a-\nu-\frac{\rm i}{2}\bigg)G_e\big(\nu,\check\la_j\big){\rm d}\nu\nonumber
\\
&\phantom{=}-2\pi {\rm i}\sum_{b=1}^{n_h}
{K\bigg(\wdp*_a-\hle_b-\frac{\rm i}{2}\bigg)}
\frac{G_e\big(\hle_b,\check\la_j\big)}{\aux'_e(\hle_b)}.
\label{intsys2_wp-}
\end{align}
Equations \eqref{intsys2_rl}, \eqref{intsys2_cpdiff}, \eqref{intsys2_cpsum},~\eqref{intsys2_wp+} and~\eqref{intsys2_wp-} describes an intricately coupled system of linear integral equation for the function $G_e$ and a linear system of equations for the remaining matrix elements for close-pairs and wide-pair.
Note that these two subsystems are highly coupled with each other too.
Our goal is to first solve the integral equation for the function~$G_e$, giving us the columns corresponding to real excited state roots.
Computing the convolutions of its solution would then give us a separated finite system of equations for the columns that corresponds to the complex roots.
This finite subsystem for complex roots emerges as a higher-level version of the original system of linear equations for the Gaudin extraction.
This result will be proved in the upcoming section, with a certain assumption on the hole spectral parameters.\looseness=1

\subsection{Emergent subsystem involving the Gaudin extraction of higher level}
\label{sub:hl}
The integral equation \eqref{intsys2_rl} can be solved to obtain
\begin{align}
 G_e\big(\rl,\check\la_j\big)={}&
 \big(1+\aux_e\big(\check\la_j\big)\big)\den_{\frac12}\bigg(\rl,\check\la_j-\frac{\rm i}{2}-{\rm i}0\bigg)
 -2\pi {\rm i}\sum_{a=1}^{n_h}\den_1(\rl,\hle_a)\frac{G_e\big(\hle_a,\check\la_j\big)}{\aux_e'(\hle_a)}\nonumber
 \\
 &+2\pi {\rm i}\sum_{a=1}^{n_{\clpx}}\big\lbrace
 \den_1(\rl,\clp^+_a)+\den_1(\rl,\clp^-_a) \big\rbrace\Fmat^{\clpx -}_{e;j,a}
 +2\pi {\rm i}\sum_{a=1}^{n_{\wdpx}}\den_1(\rl,\wdp^+_a)\Fmat^{\wdpx +}_{e;j,a}\nonumber
 \\
 &+2\pi {\rm i}\sum_{a=1}^{n_{\wdpx}}\den_1(\rl,\wdp*^-_a)\Fmat^{\wdpx -}_{e;j,a},
 \label{inteq2_rl_sol_den}
\end{align}
where the function $\den_a$ satisfy the integral equation
\begin{gather}
 \den_a(\la,\mu)+\int_{\Rset}K(\la-\nu)\den_a(\nu,\mu){\rm d}\nu=K_a(\la-\mu).
 \label{gen_liebeq}
\end{gather}
The exact form of the solutions for each of the terms in equation~\eqref{inteq2_rl_sol_den} is found in the Appendix~\ref{app:inteq}. The specific combination that is obtained for close-pairs as a result of equation~\eqref{mat2_clp_diff} leads to the conclusion that all the coefficient terms for complex roots have the exact same form, which is given by the density term for the complex roots
\begin{gather*}
 \den*(\la)=
 \frac{1}{2\pi}\frac{1}{\la^2+\frac{1}{4}} .
\end{gather*}
Therefore it would seem appropriate to construct a common combined block $\Fmat*$ corresponding to the complex roots as
\begin{subequations}\label{mat2hl}
\begin{gather}
 \Fmat*_{j,a}=\Fmat^{\clpx -}_{e;j,a},
 \label{mat2hl_cp}
 \\
 \Fmat*_{j,n_\clpx +a}=\Fmat^{\wdpx +}_{e;j,a},
 \label{mat2hl_wp+}
 \\
 \Fmat*_{j,n_\clpx+n_\wdpx+a}=\Fmat^{\wdpx -}_{e;j,a}.
 \label{mat2hl_wp-}
\end{gather}
\end{subequations}%
Equation~\eqref{inteq2_rl_sol_den} can be rewritten in terms of it as
\begin{align}
 G_e\big(\rl,\check\la_j\big)=
 \frac{\pi\big(1+\aux_e\big(\check\la_j\big)\big)}{\sinh\pi\big(\check\la_j-\rl\big)}
 -2\pi {\rm i}\sum_{a=1}^{n_h}\den_h(\rl-\hle_a)
 \frac{G_e\big(\hle_a,\check\la_j\big)}{\aux_e'(\hle_a)}
 +2\pi {\rm i}\sum_{a=1}^{\ho{n}}
 \den*(\rl-\ho{\mu}_a)
 \Fmat*_{j,a}
 \label{inteq2_rl_sol} .
\end{align}
This solution will give us first $n_r$ columns of the matrix $\Fmat_e$ that corresponds to the real roots of the excited state. The next block is made up of derived columns for close-pairs $\Fmat^{\clpx \Delta}_e$ obtained in equation~\eqref{mat2_clp_diff}.
The convolution integral can be reduced to the individual components of $G_e$ given in equation~\eqref{inteq2_rl_sol} above, leading us to the following expression
\begin{align}
 \int_{\Rset+{\rm i}0}K(\clp^+_a-\nu)G_e\big(\nu,\check\la_j\big){\rm d}\nu
 ={}& \pi\big(1+\aux_e\big(\check\la_j\big)\big)
 \int_{\Rset+{\rm i}0}K(\clp^+_a-\nu)\den_{\frac{1}{2}}\bigg(\nu,\check\la_j+\frac{\rm i}{2}\bigg){\rm d}\nu\nonumber
 \\
& -2\pi {\rm i} \sum_{a=1}^{n_h}
 \bigg\lbrace\int_{\Rset}K(\clp^+_a-\nu)\den_1(\nu,\hle_a){\rm d}\nu\bigg\rbrace\frac{G_e\big(\hle_a,\check\la_j\big)}{\aux_e'(\hle_a)}\nonumber
 \\
& +2\pi {\rm i} \sum_{a=1}^{\ho{n}}
 \bigg\lbrace\int_{\Rset}K(\clp^+_a-\nu)\den*(\nu-\ho{\mu}_a){\rm d}\nu\bigg\rbrace
 \Fmat*_{j,a}.
 \label{conv_clp_diff_cols}
\end{align}
Let us recall that $\Fmat*_e$ \eqref{mat2hl} is an another block of columns containing close-pairs and wide-pairs, distinct from $\Fmat_{e}^{c\Delta}$, all of which are embedded in the matrix $\Fmat_{e}$.
Since the equation~\eqref{conv_clp_diff_cols} computes a part of equation~\eqref{intsys2_cpdiff} solving for $\Fmat_{\rm e}^{c\Delta}$, we immediately find out that the last term in this equation~\eqref{conv_clp_diff_cols} can be cancelled upto the determinant by subtracting linear sum over columns $\Fmat*_e$.
In other words we find that the coefficients of the last sum in equation~\eqref{conv_clp_diff_cols} are not important for the purpose of computing determinant of $\Fmat_e$. Hence we do not compute convolutions with $\den*$ in the last term and we shall treat it as free coefficients $\set{\chi_a}$.

We note that as product of rational functions $K_1(\clp^+-\nu)$ and $\den*(\nu-\ho\mu)=K_{\frac{1}{2}}(\nu-\ho\mu)$, it is an integrable function which we will write as undetermined coefficients $\chi$ from now on.
This leaves us with the convolutions of $\den_1\big(\nu,\la+\frac{\rm i}{2}\big)$ and $\den_{\frac{1}{2}}(\nu,\hle)$ with the shifted Lieb kernel $K(\clp^+-\nu)$.
As it turns out, convolution integrals involved here can be evaluated directly from the integral equation \eqref{gen_liebeq} by deforming its contour
\begin{gather*}
 \den_{\frac{1}{2}}\bigg(\clp^+_a,\la+\frac{\rm i}{2}\bigg)+
 \int_{\Rset+{\rm i}\alpha}K(\clp^+_a-\nu)\den_{\frac{1}{2}}\bigg(\nu,\la+\frac{\rm i}{2}\bigg){\rm d}\nu =\frac{1}{2\pi {\rm i}}t(\la-\clp^+_a),
 \\
 \den_1(\clp^+_a,\hle_b)+\int_{\Rset+{\rm i}\alpha}K(\clp^+_a-\nu)\den_1(\nu,\hle_b){\rm d}\nu
 =K_1(\clp^+_a-\hle_b).
\end{gather*}
This gives us a very similar expression
\begin{gather}
 \Fmat^{\clpx \Delta}_{e;j,a}=
 \frac{\pi}{\sinh\pi\big(\check\la_j-\clp_a-\frac{\rm i}{2}\big)}
 -2\pi {\rm i} \sum_{a=1}^{n_h}\den_h\bigg(\clp_a+\frac{\rm i}{2}-\hle_b\bigg)
 \frac{G_e\big(\hle_b,\check\la_j\big)}{\aux_e'(\hle_b)}
 +2\pi {\rm i}\sum_{b=1}^{\ho{n}}\chi_b\Fmat*_{j,b}.
 \label{mat2_clp_sol}
\end{gather}

The remaining block consists of higher-level columns $\Fmat*$ block as introduced in equation~\eqref{mat2hl}. Previously, we have obtained equations~\eqref{intsys2_cpsum}, \eqref{intsys2_wp+} and~\eqref{intsys2_wp-} for elements of these columns.
Now the remaining convolutions integrals involved in these equations can be computed after substituting the expression \eqref{inteq2_rl_sol} for $G_e$:
\begin{gather*}
 \int_{\Rset+{\rm i}\alpha}
 \begin{pmatrix}
 K^{(c)}(\clp-\nu) \\ K(\wdp^+-\nu) \\ K(\wdp*^--\nu)
 \end{pmatrix}
 G_e\big(\nu,\check\la_j\big){\rm d}\nu
 \\ \qquad
{} = \pi\big(1+\aux_e\big(\check\la_j\big)\big)\int_{\Rset+{\rm i}\alpha}
\begin{pmatrix}
 K^{(c)}(\clp-\nu) \\ K(\wdp^+-\nu) \\ K(\wdp*^--\nu)
 \end{pmatrix}
 \den_{\frac{1}{2}}\bigg(\nu,\check\la_j+\frac{\rm i}{2}\bigg){\rm d}\nu
 \\ \qquad\phantom{=}
 {}-2\pi {\rm i}\sum_{a=1}^{n_h}
 \left\lbrace \int_{\Rset+{\rm i}\alpha}
 \begin{pmatrix}
 K^{(c)}(\clp-\nu) \\ K(\wdp^+-\nu) \\ K(\wdp*^--\nu)
 \end{pmatrix}
 \den_1(\nu,\hle_a){\rm d}\nu
\right\rbrace\frac{G_e\big(\check\la_j,\hle_a\big)}{\aux_e(\hle_a)}
 \\ \qquad\phantom{=}
 {}+2\pi {\rm i}\sum_{a=1}^{\ho n}
 \left\lbrace
 \int_{\Rset+{\rm i}\alpha}
 \begin{pmatrix}
 K^{(c)}(\clp-\nu) \\ K(\wdp^+-\nu) \\ K(\wdp*^--\nu)
 \end{pmatrix}
 \den*(\nu-\ho\mu_a){\rm d}\nu\right\rbrace\Fmat*_{j,a}.
\end{gather*}
The convolution integrals thus obtained, unlike in the case of similar convolutions with shifted kernel $K^{(c)}$ or $K$, needs to be explicitly calculated.
The explicit results for these calculations are presented in Appendix~\ref{app:den_conv}.
By substituting these results, we obtain a finite linear system of equations for $\Fmat*$ elements. We can also check that it takes a similar form in all three cases corresponding to the close-pairs and wide-pairs, this combined system can be written as
\begin{gather}
 \aux*'(\ho{\mu}_a)\Fmat*_{j,a}
 -2\pi {\rm i}\sum_{b=1}^{\ho{n}}K(\ho{\mu}_a-\ho{\mu}_b)\Fmat*_{j,b}
 = -2\pi {\rm i}\sum_{b=1}^{n_h}
 \den*(\ho{\mu}_a-\hle_b) \frac{G_e\big(\hle_b,\check\la_j\big)}{\aux_e'(\hle_b)} .
 \label{hlgauex_linsys_coupled}
\end{gather}
This subsystem bears a striking resemblance with the original system of linear equations for the Gaudin extraction \eqref{linsys}.
The left hand side can be interpreted as the action of inverse of the \emph{higher-level} Gaudin matrix
\begin{gather}
 \Gau*_{a,b}=
 \aux*'(\ho{\mu}_a)\delta_{a,b}-2\pi {\rm i}\,K(\ho{\mu}_a-\ho{\mu}_b).
 \label{HLGau}
\end{gather}
On the right-hand side of equation~\eqref{hlgauex_linsys_coupled} we have the matrix formed by density function $\den*^{h}$ for complex roots
\begin{gather}
 \Den*^{h}_{a,b}=-2\pi {\rm i}\,\den*(\ho{\mu}_a-\hle_b).
 \label{hldenmat}
\end{gather}
In terms of it we can write equation~\eqref{hlgauex_linsys_coupled} in the following way
\begin{gather}
 \Fmat*= \Hmat \bigg[ \frac{G_e\big(\hle_a,\check\la_j\big)}{\aux'_e(\hle_a)} \bigg]_{a\leq n_h},
 \label{mat2hl_sol}
\end{gather}
where the matrix $\Hmat$ solves the higher-level equivalent of the Gaudin extraction
\begin{gather}
 \Hmat=\Gau*^{-1}\Den*^h .
 \label{HLmat}
\end{gather}
Substituting equation~\eqref{mat2hl} into equation~\eqref{inteq2_rl_sol} gives us an auxiliary system for the matrix composed of $G_e\big(\hle_a,\check\la_j\big)$.
It can be written in the matrix form as
\begin{gather}
 \big(\Id_{n_h} -\Den^{\text{dr}}(\Aux_e')^{-1}\big)
 \big[G_e\big(\hle_a,\check\la_j\big)\big]_{a\leq n_h}
 = \bigg[\frac{1}{\sinh\pi\big(\check\la_j-\hle_a\big)}\bigg]_{a\leq n_h}.
 \label{mat2_auxsys}
\end{gather}
The dressed matrix $\Den^{\text{dr}}$ is given by the expression
\begin{gather*}
 \Den^{\text{dr}}= \Den_h-\big(\Den*^h\big)^T\Gau*\Den*^h
\end{gather*}
in which the matrix $\Den_h$ is defined in similar manner to the matrix $\Den*^h$ \eqref{hldenmat} as
\begin{gather*}
 \Den_{h;a,b}=
 -2\pi {\rm i}\den_h(\hle_a-\hle_b)
\end{gather*}
whereas the matrix $\Aux'_e$ is the diagonal matrix composed of
\begin{gather*}
 \Aux'_{e;a,b}=\aux_e'(\hle_a)\delta_{a,b}.
\end{gather*}
Let us note that when the hole parameters $\set{\hle_a}_{a\leq n_h}$ are chosen within the bulk of the Fermi distribution, all the diagonal elements of this matrix scale as $\Aux'_{e;a,a}\sim M$.
Hence the system equation~\eqref{mat2_auxsys} completely decouples to give the following expression upto leading order
\begin{gather*}
 G_e\big(\hle_a,\check\la_j\big)=
 \frac{1}{\sinh\pi\big(\check\la_j-\hle_a\big)} + O\big(M^{-1}\big).
\end{gather*}
Substituting this decoupled expression into equations~\eqref{inteq2_rl_sol}, \eqref{mat2_clp_sol} and~\eqref{mat2hl_sol} allows us to write the elements of $\Fmat_e$ as follows.
Taking a cue from the similarity of the expressions~\eqref{inteq2_rl_sol} and~\eqref{mat2_clp_sol}, we will write the blocks $\Fmat^{\rlx}_e$ and $\Fmat^{\clpx\Delta}_e$ as
\begin{gather}
 \Fmat^{\rcx}_{e;j,k}=
 \frac{\pi\big(1+\aux_e\big(\check\la_j\big)\big)}{\aux'_e(\rl_k)}
 \Bigg\lbrace \frac{1}{\sinh\pi\big(\check\la_j-\rl_k\big)}
 -2\pi {\rm i}\sum_{b=1}^{n_h}
 \frac{\den_h(\rl_k-\hle_b)}{\aux'_e(\hle_b)}\frac{1}{\sinh\pi\big(\check\la_j-\hle_b\big)}
 \Bigg\rbrace,
 \label{mat2_cau_rl}
 \\
 \Fmat^{\rcx}_{e;j,n_\rlx+a}= {\pi\big(1+\aux_e\big(\check\la_j\big)\big)}\nonumber
 \\ \hphantom{ \Fmat^{\rcx}_{e;j,n_\rlx+a}=}
 {}\times \Bigg\lbrace
 \frac{1}{\sinh\pi\big(\check\la_j-\clp_a-\frac{\rm i}{2}\big)} -2\pi {\rm i}\sum_{b=1}^{n_h}
 \frac{\den_h\big(\clp_a+\frac{\rm i}{2}-\hle_b\big)}{\aux'_e(\hle_b)}\frac{1}{\sinh\pi\big(\check\la_j-\hle_b\big)}
 \Bigg\rbrace.
 \label{mat2_cau_cp}
\end{gather}
This forms an infinite sub-block of $\Fmat_e$ which can be written as sum over columns of a larger Cauchy matrix that can be constructed using all real roots including holes.
Notice that we have cancelled the linear sum over columns of the matrix $\Fmat*$ sub-block in equations~\eqref{mat2hl_sol} and~\eqref{mat2_clp_sol} while writing above equations~\eqref{mat2_cau_cp} and~\eqref{mat2_cau_rl} for $\Fmat^{\rcx}_e$ since these terms are automatically cancelled in the determinant of the matrix $\Fmat_e$.

The elements of the remaining columns can be found from\vspace{-1ex}
\begin{gather}
 \Fmat*_{j,a}=
 \pi\big(1+\aux_e\big(\check\la_j\big)\big)\sum_{b=1}^{n_h}\Hmat_{a,b}\frac{1}{\sinh\pi\big(\check\la_j-\hle_b\big)},\qquad
 \Hmat = \Gau*^{-1}\Den*^h .
 \label{mat2HL}
\end{gather}
Equations~\eqref{mat2_cau_rl}--\eqref{mat2HL} collectively describe the matrix $\Fmat_e$ in the presence of bound state. Let us stress on the importance of the result \eqref{mat2HL}. We have found a finite linear subsystem~\eqref{hlgauex_linsys_coupled} to describe this block of columns. We also show that it can be interpreted as an auxiliary system~\eqref{HLmat} arising from the extraction of higher-level Gaudin matrix $\Gau*$ \eqref{HLGau}.

\section[Form factors as determinants of perturbed Cauchy matrices]
{Form factors as determinants of perturbed \\Cauchy matrices}

We have previously shown that ratio of determinants can be replaced by a single determinant using Gaudin extraction. In this paper we have taken this approach further as we considered form factors for excitations containing complex Bethe roots in the form of 2-string, quartets or wide-pairs. The presence of such complex numbers is the main source of perturbation to a predominantly Cauchy matrix structure obtained from Gaudin extraction with real Bethe roots.
Inserting all the results obtained in this paper into \eqref{det_rep_finite}, we obtain the following determinant representation for form factors\vspace{-1ex}
\begin{gather}
 |\mathcal{F}_z|^2=-2
 \prod_{a=1}^{\frac{M}{2}-1}\frac{q_g(\mu_a-{\rm i})}{\bar q_e(\mu_a-{\rm i})}
 \prod_{a=1}^{\frac{M}{2}}\frac{q_e(\la_a-{\rm i})}{q_g(\la_a-{\rm i})}
 \det_{\frac{M}{2}}\Fmat_g\det_{\frac{M}{2}+1}\Fmat_e.
 \label{fin_det_rep}
\end{gather}
Functions $q_g$ and $q_e$ in the prefactor of equation~\eqref{fin_det_rep} are Baxter polynomials for the ground state and the excited state respectively. It is defined as the product \eqref{baxq}. The $\bar q_e$ in the denominator is defined identically to $q_e$ except for close-pair roots where $\bar q_e$ is regularised as follows
\begin{gather*}
 \bar q_e(\clp_a^++{\rm i}\sdv_a-{\rm i})=
 \frac{1}{2{\rm i}\sdv_a}\, q_e(\clp^+_a+{\rm i}\sdv_a-{\rm i}).
 \end{gather*}
In this way it absorbs the factors $(2{\rm i}\sdv_a)$ previously extracted from the determinant of $\Fmat_e$ into the prefactors by the virtue of equation~\eqref{mat2_clp_diff}.
The matrices $\Fmat_g$ and $\Fmat_e$ in equation~\eqref{fin_det_rep} are perturbed Cauchy matrices obtained from the Gaudin extraction.
Columns of these matrices are as follows:
\begin{gather*}
\begin{split}
& \Fmat_g=
 \big(\Fmat_g^{rc+}\mid\Fmat_g^{c-}\mid\Fmat_{g}^{w+}\mid\Fmat_g^{w-}\big),
 \\
 &\Fmat_e=
 \big(\Fmat_{\rm e}^{rc+}\mid\Ho{\Fmat}_e\mid\FW\big).
\end{split}
\end{gather*}
It contains large Cauchy submatrices $\Fmat_g^{rc+}$ and $\Fmat_{\rm e}^{rc-}$ in both cases:
\begin{subequations}
\label{fin_Fmat_g_cau}
\begin{gather}
 \Fmat_{g;j,k}^{\rcx}= \frac{1+\aux(\check\rl_k)}{\aux_g'(\la_j)} \frac{\pi}{\sinh\pi(\check\rl_k-\la_j)},
 \\
 \Fmat_{g;j,{n_r+a+1}}^{\rcx}= \frac{1}{\aux_g'(\la_j)} \frac{\pi}{\sinh\pi(\clp^+_a-\la_j)},
\end{gather}
\end{subequations}
\vspace{-\baselineskip}
\begin{subequations}
\label{fin_fmat_e_cau}
\begin{gather}
 \Fmat^{\rcx}_{e;j,k}= \frac{\pi\big(1+\aux_e\big(\check\la_j\big)\big)}{\aux'_e(\rl_k)}
 \Bigg\lbrace \frac{1}{\sinh\pi\big(\check\la_j-\rl_k\big)} -2\pi {\rm i}\sum_{b=1}^{n_h}
 \frac{\den_h(\rl_k-\hle_b)}{\aux'_e(\hle_b)}\frac{1}{\sinh\pi\big(\check\la_j-\hle_b\big)}\Bigg\rbrace,
 \label{fin_mat2_cau_rl}
 \\
 \Fmat^{\rcx}_{e;j,n_\rlx+a}= {\pi\big(1+\aux_e\big(\check\la_j\big)\big)}\nonumber
 \\ \hphantom{\Fmat^{\rcx}_{e;j,n_\rlx+a}=}
 {}\times\Bigg\lbrace \frac{1}{\sinh\pi\big(\check\la_j-\clp_a-\frac{\rm i}{2}\big)} -2\pi {\rm i}\sum_{b=1}^{n_h}
 \frac{\den_h\big(\clp_a+\frac{\rm i}{2}-\hle_b\big)}{\aux'_e(\hle_b)}\frac{1}{\sinh\pi\big(\check\la_j-\hle_b\big)}
 \Bigg\rbrace.
 \label{fin_mat2_cau_cp}
\end{gather}
\end{subequations}
Note that this also includes a partial contribution of close-pairs (whenever it is present) since we have seen how close-pair contribution splits into two halves where one part behaves as if it was a contribution from a real root. The non-trivial perturbation to this presentation contains a~mixture of other half of the close-pair columns and wide-pair columns.
In the matrix $\Fmat_g$, these perturbed columns due to complex roots takes the form
\begin{gather*}
 \Fmat^{c-}_{g;j,a}=
 \frac{\aux_g(\clp^-_a-{\rm i}\sdv_a)}{\aux_g'(\la_j)}
 \bigg\lbrace \frac{\pi}{\sinh\pi(\clp^+_a-\la_j+{\rm i}\sdv_a)}
 + \frac{\pi}{\sinh\pi(\clp^-_a-\la_j-{\rm i}\sdv_a)} \bigg\rbrace\nonumber
 \\ \hphantom{ \Fmat^{c-}_{g;j,a}=}
 +\frac{2\pi {\rm i}}{\aux_g'(\la_j)}
\big\lbrace \den_{1}(\la_j,\clp^+_a)-\den_{1}(\la_j,\clp^-_a) \big\rbrace,
\\
 \Fmat_{g;j,a}^{\wdpx+}=
 \frac{2\pi {\rm i}}{\aux_g'(\la_j)} \big\lbrace
 \den_{\frac12}(\la_j,\wdp_a+{\rm i})-\den_{\frac12}(\la_j,\wdp_a) \big\rbrace,
 \\
 \Fmat_{g;j,a}^{\wdpx-}=
 \frac{2\pi {\rm i}}{\aux_g'(\la_j)} \big\lbrace
 \den_{\frac12}(\la_j,\wdp*_a)-\den_{\frac12}(\la_j,\wdp*_a-{\rm i}) \big\rbrace.
\end{gather*}
Whereas the perturbation due to complex roots in the matrix $\Fmat_e$ takes a more unified form, where a block of columns $\Fmat*_e$ unifying remaining pair columns and wide-pair columns takes shape. Moreover it can be expressed using a Cauchy convolution relation
\begin{gather*}
 \Fmat*_{j,a}=
 \pi\big(1+\aux_e\big(\check\la_j\big)\big)\sum_{b=1}^{n_h}\Hmat_{a,b}\frac{1}{\sinh\pi\big(\check\la_j-\hle_b\big)}.
\end{gather*}
We say that the kernel of this convolution $\Hmat$ is a result of higher-level Gaudin extraction because it satisfies
\begin{gather}
 \Hmat = \Gau*^{-1}\Den*^h,
 \label{fin_hlgauex}
\end{gather}
where the matrix $\Gau*$ is a Gaudin matrix corresponding to the higher-level Bethe equations \eqref{hlbe}
\begin{gather*}
 \Gau*_{a,b}=
 \aux*'(\ho\mu_a)\delta_{a,b}-2\pi {\rm i} K(\ho\mu_a-\ho\mu_b).
\end{gather*}
Let us recall that $\Den*^h$ is a matrix of complex density function \eqref{hldenmat} as shown below
\begin{gather*}
 \Den*^h_{a,b}=-2\pi {\rm i}\den*(\hle_a-\ho\mu_b)=-2\pi {\rm i}K_{\frac 12}(\hle_a-\ho\mu_b).
\end{gather*}
In addition to the columns $\Fmat*_e$, we also have a trivial perturbation due to Foda--Wheeler columns~$\FW$ present in~$\Fmat_e$. It does not participate in Gaudin extraction and hence retains its original form
\begin{gather*}
 \FW_{j,1}=\aux_g\big(\check\la_j\big)-1,\qquad
 \FW_{j,2}=\aux_g\big(\check\la_j\big)\big(\check\la_j+{\rm i}\big)-\la_j.
\end{gather*}
Finally we would like to point out that the determinant representation \eqref{fin_det_rep} can be further simplified. Similar to our approach in~\cite{KitK19}, we can show that the determinants of dominant Cauchy matrices can be factored out using the ``Cauchy extraction''~\cite{Kul20th}.
Briefly, this process includes taking an action of an inverse matrix in such a way that the action on largest Cauchy block present in $\Fmat_g$ and $\Fmat_e$ is trivial.
Whereas result of its action on the columns of finite rank perturbation (non-Cauchy columns) is non-trivial. To the level of their determinants, the resulting matrices can be reduced to finite matrices $\mathcal{Q}_g$ and $\mathcal{Q}_e$ whose elements can be written as explicit integrals containing an auxiliary function jointly determined by the excited state and ground state.
This allows us to write down the following formula
\begin{gather}
 |\mathcal{F}_z|^2=
 M^{-n_h}S(\set{\hle_a}_{a=1}^{n_h})
 \frac{\prod_{a=1}^{\ho n}\prod_{b=1}^{n_h}\big(\ho\mu_a-\hle_b-\frac{\rm i}{2}\big)}{\prod_{a,b}^{\ho n}(\ho\mu_a-\ho\mu_b-{\rm i})\prod_{a<b}^{n_h}(\hle_a-\hle_b)}
 {\det_{\ho n}\mathcal{Q}_g\det_{n_h}\mathcal{Q}_e}.
 \label{bnd_st_det_rep}
\end{gather}
The function $S$ in equation~\eqref{bnd_st_det_rep} is the spinon scattering factor that is obtained from the thermodynamic limit of the determinants of the Cauchy matrices which we had extracted. It is can be expressed in terms of the Barnes G-function
\begin{gather*}
 S(\set{\hle_a}_{a=1}^{n_h})= (-1)^{\frac{n_h+2}{2}}
 2^{\frac{n_h(n_h-2)+2}{2}} \pi^{\frac{n_h(n_h-3)+2}{2}}
\nonumber
 \\ \hphantom{ S(\set{\hle_a}_{a=1}^{n_h})= }{}\times\frac{1}{G^{2n_h}\big(\frac{1}{2}\big)} \prod_{a\neq b}^{n_h}
\frac{G\big(\frac{\hle_a-\hle_b}{2{\rm i}}\big)G\big(1+\frac{\hle_a-\hle_b}{2{\rm i}}\big)}
 {G\big(\frac{1}{2}+\frac{\hle_a-\hle_b}{2{\rm i}}\big)G\big(\frac{3}{2}+\frac{\hle_a-\hle_b}{2{\rm i}}\big)}.
\end{gather*}
The detailed derivation of this formula as well as explicit form of finite matrices $\mathcal{Q}_g$ and $\mathcal{Q}_e$ will be given in a forthcoming publication. However
it is worthwhile to remark that $\Fmat*_e$ is expressed as sums of Cauchy terms in hole rapidities with coefficients given by $\Hmat$. The action of an inverse Cauchy matrix can be taken in such a way
 that the higher-level matrix $\Hmat$ equation~\eqref{fin_hlgauex} gets directly embedded in $\mathcal{Q}_e$ occupying $\ho{n}=\frac{n_h}{2}-1$ columns in $\mathcal{Q}_e$.

Similarly, we can see that columns of $\Fmat_{\rm e}^{\rcx}$ \eqref{fin_fmat_e_cau} include sums of Cauchy terms in hole rapidities with coefficients scaling as $M^{-1}$ since we have $\aux'_e(\hle_a)\sim M$ for holes located in the bulk. As a result of the Cauchy extraction, this summation term gets resolved into individual columns, one for each rapidity. These columns are embedded into the $\mathcal{Q}_e$ in equation~\eqref{bnd_st_det_rep} after rescaling them with a factor of $M$. This is how we obtain the power law $M^{-n_h}$ in prefactor of equation~\eqref{bnd_st_det_rep}.

The remaining elements of finite matrices $\mathcal{Q}_g$ as well as $\mathcal{Q}_e$ can be written explicitly as integrals with certain auxiliary functions. Together with the rational prefactor in equation~\eqref{bnd_st_det_rep}, they are functions of complex roots which are themselves determined implicitly from hole positions $\set{\hle_a}_{1}^{n_h}$ through the higher level Bethe equations \eqref{hlbe}. Further treatment of these objects deserves special attention and it will be dealt in a future publication. We will also provide the details of computations involved in the Cauchy extraction that allowed us to write the determinant representation \eqref{bnd_st_det_rep} starting from equation~\eqref{fin_det_rep}.

\section{Conclusions}
\label{sec:concl}
We have shown that all types of complex Bethe roots can be taken into account in the framework of the ``Gaudin extraction'' approach introduced in our previous paper. We show that the form factors for the low-lying excited states can be computed as determinants of Cauchy matrices perturbed by a finite rank matrix (with a rank less than or equal to the number of spinons). It means that the result can be written as a Cauchy determinant easily computable in the thermodynamic limit (and leading to the product of the spinon scattering factors) and a~determinant of a finite dimensional matrix containing all the dependency on the complex roots. In particular we show how the Gaudin determinant of the higher level Bethe equations emerges in the framework of this approach.

\appendix
\section{Condensation property for meromorphic functions}
\label{app:gen_condn}
In the proposed method, we find it sometime useful to extend the reach of condensation property to include meromorphic functions \cite{KitK19}. For this purpose let us consider a meromorphic function~$g$ with a set of simple poles $\{z_{a}\}_{a\leq n}$ non overlapping with Bethe roots $\{z_a\}\cap\set{\la_a}=\varnothing$.
Since we have,
\begin{gather*}
 g(\la_a)= \res\limits_{z=\la_a}\frac{-2\pi {\rm i}M \den(z)}{1+\aux(z)}g(z)
\end{gather*}
we can convert the sum
\begin{gather*}
 \frac{1}{M}\sum_{a=1}^{N}g(\la_a)=
 \bigg(\int_{\Rset+{\rm i}\alpha}-\int_{\Rset-{\rm i}\alpha}\bigg)g(\la)\frac{\den(\la)}{1+\aux(\la)}{\rm d}\la
 +2\pi {\rm i}\sum_{a=1}^{n}\frac{\den(z_a)}{1+\aux(z_a)}\res g(z_a).
\end{gather*}
It is necessary to have the function $g$ diminishing sufficiently rapidly at the tails, so that the edges of the contour can be ignored.
We also assume that poles of the function $g$ are within the bulk of distribution where we can write
\begin{gather*}
 \aux(\nu)=O\big({\rm e}^{-{\rm sign}(\operatorname{Im}\nu)M}\big).
\end{gather*}
This would allow us to convert the sum over Bethe roots for a meromorphic function to a density integrals with extra residue terms
\begin{gather*}
 \frac{1}{M}\sum_{a=1}^{N}g(\la_a)=
 \int_{\Rset+{\rm i}\alpha}g(\la)\den(\la){\rm d}\la+
 2\pi {\rm i}\sum_{a=1}^{n}\frac{\den(z_a)}{1+\aux(z_a)}\res g(z_a).
\end{gather*}
We refer to this as generalised condensation property.

\section{Integral equations for generalised density functions}
\label{app:inteq}
Let us first consider a generalised form of integral equation
\begin{gather*}
 \den_a(\la,\mu)+\int_{\Rset}K(\la-\nu)\den_a(\nu,\mu){\rm d}\nu = K_a(\la-\mu).
\end{gather*}
The solution of this integral equation depends on the parameter $\mu\in\Cset$ and scaling parameter $a\in\Rset$. Since $K_a(\la-\mu)$ has poles at $\mu\pm ia$, we find that its Fourier transform has branches
\begin{gather*}
 \widehat{\den}_{a}(t,\mu)=
 \begin{dcases}
 \frac{{\rm e}^{-a|t|-{\rm i}\mu t}}{1+{\rm e}^{-t}}, & |\operatorname{Im}\mu|<a,
 \\
 I_{\sigma t>0} \frac{\sinh(\sigma a t)\, {\rm e}^{-{\rm i}\mu t}}{1+{\rm e}^{-\sigma t}}, &
 \sigma\operatorname{Im}\mu>a,\quad \sigma=\pm 1.
 \end{dcases}
\end{gather*}
In particular, we are interested in the cases $a=\frac{1}{2}$ and $a=1$. We know that the $\den_{\frac12}(\la,\mu)$ represents the generalisation of Lieb equation for the ground state density $\den_g$.
We find that this solution can be extended to the entire region $|\operatorname{Im}\mu|<\frac{1}{2}$
\begin{gather}
 \den_\frac{1}{2}(\la,\mu)= \den_g(\la-\mu)
 = \frac{1}{2\cosh\pi(\la-\mu)},\qquad
 \text{inside }|\operatorname{Im}\mu|<\frac{1}{2}.
 \label{gen_den_cau}
\end{gather}
Note that this also applies to the limiting case $\den_\frac{1}{2}\big(\la,\frac{\rm i}{2}-{\rm i}0\big)$ and it translates to the solution of deformed integral equation
\begin{gather*}
 \den_\frac{1}{2}\bigg(\la,\frac{\rm i}{2}-{\rm i}0\bigg)
 +\int_{\Rset+{\rm i}0}K(\la-\nu)\den_{\frac{1}{2}}\bigg(\nu,\frac{\rm i}{2}-{\rm i}0\bigg){\rm d}\nu
 = t\bigg(\la-\frac{\rm i}{2}\bigg),
\end{gather*}
which we frequently encounter in our computations. Hence the extensibility of solution for $\den_g$ in the form equation~\eqref{gen_den_cau} means that we obtain the Cauchy matrix.

However in the outside $|\operatorname{Im}\mu|>\frac12$ we get different branch of the solution to the integral equation
\begin{gather*}
 \den_{\frac12}(\la,\mu)= \frac{1}{4\pi}\bigg\lbrace
 \Psi\bigg(\frac12-\frac{\la-\mu}{2{\rm i}\sigma}\bigg)-\Psi\bigg({-}\frac{\la-\mu}{2{\rm i}\sigma}\bigg)
 \bigg\rbrace,\qquad
 \sigma\operatorname{Im}\mu>\frac12 .
\end{gather*}

The solution $\den_{1}(\la,\mu)$ inside $|\operatorname{Im}\mu|<1$ is obtained as
\begin{gather*}
\den_1(\la,\mu)= \frac{1}{4\pi}\bigg\lbrace
\Psi\bigg(1+\frac{\la-\mu}{2{\rm i}}\bigg) + \Psi\bigg(1-\frac{\la-\mu}{2{\rm i}}\bigg)
- \Psi\bigg(\frac12-\frac{\la-\mu}{2{\rm i}}\bigg) \nonumber
\\ \hphantom{\den_1(\la,\mu)= \frac{1}{4\pi}\bigg\lbrace}
{}- \Psi\bigg(\frac12-\frac{\la-\mu}{2{\rm i}}\bigg)\bigg\rbrace,
\end{gather*}
while outside in $|\operatorname{Im}\mu|>1$ we get
\begin{gather*}
 \den_1(\la,\mu)=\frac{1}{2\pi {\rm i}}\frac{1}{(\la-\mu)(\la-\mu+{\rm i}\sigma)}=K_{\frac12}\bigg(\la-\mu-\frac{{\rm i}\sigma}{2}\bigg),\qquad
 \sigma\operatorname{Im}\mu>1.
\end{gather*}
Note that in particular we have the factorisation identity that can be invoked for the close-pairs
\begin{gather*}
 \den_1\bigg(\la,\mu-\frac{\rm i}{2}\bigg)+\den_1\bigg(\la,\mu+\frac{\rm i}{2}\bigg)=
 \frac{1}{2\pi {\rm i}} \frac{1}{(\la-\mu)^2+\frac{1}{4}}=
 K_{\frac12}(\la-\mu), \qquad
 |\operatorname{Im}\mu|<\frac{1}{2}.
\end{gather*}

\section{Density convolutions}
\label{app:den_conv}
In this appendix we give the results of convolution integrals that were encountered in Section~\ref{sub:hl}.
We will tabulate them into different categories according to its kernel and density function involved in it.

\subsection*{Density convolutions with $\boldsymbol{K^{(c)}}$}
For this part it is important to note that the Fourier transform of the kernel $K^{(c)}(\la)=K\big(\la\allowbreak+\frac{\rm i}{2}\big)+K\big(\la-\frac{\rm i}{2}\big)$ takes a simple form
\begin{gather*}
 \widehat{K^{(c)}}(t)={\rm e}^{-\frac{|t|}{2}}\big(1+{\rm e}^{-|t|}\big).
\end{gather*}

\medskip\noindent{\it With the density term $\den_{\frac{1}{2}}$:}
\begin{gather*}
 \int_{\Rset}K^{(c)}(\clp_a-\nu)\den_{\frac{1}{2}}\bigg(\nu,\la+\frac{\rm i}{2}\bigg){\rm d}\nu
 =
\frac{1}{2\pi}\int_{\Rset^+}{\rm e}^{{\rm i}\left(\clp_a-\la-\frac{\rm i}{2}\right)t-|t|}{\rm d}t
 =
 K(\clp^+_a-\la).
\end{gather*}

\medskip\noindent{\it With the density term $\den_1$:}
\begin{gather*}
 \int_{\Rset}K^{(c)}(\clp_a-\nu)\den_1(\nu,\hle_b){\rm d}\nu=
 \frac{1}{2\pi}\int_{\Rset^+}{\rm e}^{{\rm i}(\clp_a-\hle_b)t-\frac{3|t|}{2}}{\rm d}t=K_{\frac{3}{2}}(\clp_a-\hle_b).
\end{gather*}

\medskip\noindent{\it With the density term $\den*$:}
For the density of complex roots, it is necessary to distinguish between close-pairs and wide-pairs. We can write the results in the three distinct scenarios as
\begin{gather*}
 \int_{\Rset}K^{(c)}(\clp_a-\nu)
 \begin{pmatrix}
 \den*(\nu-\clp_b) \\ \den*(\nu-\wdp_b) \\ \den*(\nu-\wdp*_b)
 \end{pmatrix}
 {\rm d}\nu =
 \begin{pmatrix}
 K(\clp_a-\clp_b)-K_2(\clp_a-\clp_b) \\ K(\clp_a-\wdp_b-{\rm i})\\
 K(\clp_a-\wdp*_b+{\rm i})
 \end{pmatrix}\!.
\end{gather*}

\subsection*{Density convolutions for $\boldsymbol K$ shifted by wide-pairs}

\noindent{\it With the density term $\den_{\frac{1}{2}}$:}
\begin{align*}
 \int_{\Rset}
 \begin{pmatrix}
 K(\wdp^+_a-\nu) \\ K(\wdp^-_a-\nu)
 \end{pmatrix}
 \den_{\frac{1}{2}}\bigg(\nu,\la+\frac{\rm i}{2}\bigg){\rm d}\nu =
 \frac{1}{\pi}\int_{\Rset^+}\sinh\frac{t}{2}
 \begin{pmatrix}
 {\rm e}^{{\rm i}\wdp_at} \\ {\rm e}^{-{\rm i}\wdp*_at}
 \end{pmatrix}{\rm e}^{-\frac{t}{2}} {\rm d}t = \frac{1}{2\pi {\rm i}}
 \begin{pmatrix}
 K_{\frac{1}{2}}(\wdp_a) \\ K_{\frac{1}{2}}(\wdp*_a-{\rm i})
 \end{pmatrix}\!.
\end{align*}

\medskip\noindent{\it With the density term $\den_{1}$:}
\begin{align*}
 \int_{\Rset}
 \begin{pmatrix}
 K(\wdp^+_a-\nu) \\ K(\wdp*^-_a-\nu)
 \end{pmatrix}
 \den_1(\nu,\hle_b){\rm d}\nu =
 \frac{1}{\pi}\int_{\Rset^+} \sinh\frac{t}{2}
 \begin{pmatrix}
 {\rm e}^{{\rm i}(\wdp_a-\hle_b)} \\ {\rm e}^{-{\rm i}(\wdp*_a-\hle_b)}
 \end{pmatrix}{\rm e}^{-t}{\rm d}t =
 \begin{pmatrix}
 K_{\frac{1}{2}}(\wdp_a-\hle_b+{\rm i})
 \\
 K_{\frac{1}{2}}(\wdp*_a-\hle_b-{\rm i})
 \end{pmatrix}\!.
\end{align*}

\medskip\noindent{\it With the density term $\den*$ for close-pairs:}
\begin{align*}
 \int_{\Rset}
 \begin{pmatrix}
 K(\wdp^+_a-\nu) \\ K(\wdp*^-_a-\nu)
 \end{pmatrix}
 \den*(\nu-\clp_b){\rm d}\nu &=
 \frac{1}{\pi}\int_{\Rset^+}\sinh t
 \begin{pmatrix}
 {\rm e}^{{\rm i}(\wdp_a-\clp_b)t} \\ {\rm e}^{-{\rm i}(\wdp*_a-\clp_b)t}
 \end{pmatrix}{\rm e}^{-t}{\rm d}t= \begin{pmatrix}
 K(\wdp_a-\clp_b+{\rm i})\\ K(\wdp*_a-\clp_b-{\rm i})
 \end{pmatrix}\!.
\end{align*}

\medskip\noindent{\it With the density terms $\den*$ for wide-pairs:}
\begin{gather*}
 \int_{\Rset}K(\wdp^+_a-\nu)\den*(\nu-\wdp_b)\,{\rm d}\nu=0,
 \\
 \int_{\Rset}K(\wdp^+_a-\nu)\den*(\nu-\wdp*_b)\,{\rm d}\nu=
 \frac{1}{\pi}\int_{\Rset^+}\sinh t({\rm e}^{-t}-1){\rm e}^{{\rm i}(\wdp_a-\wdp*_b)t}{\rm d}t
 \\ \hphantom{ \int_{\Rset}K(\wdp^+_a-\nu)\den*(\nu-\wdp*_b)\,{\rm d}\nu}
 {}= K(\wdp_a-\wdp*_b+{\rm i})-K(\wdp_a-\wdp*_b),
 \\
 \int_{\Rset}K(\wdp*^-_a-\nu)\den*(\nu-\wdp_b)\,{\rm d}\nu=
 \frac{1}{\pi}\int_{\Rset^+}\sinh t({\rm e}^{-t}-1){\rm e}^{-{\rm i}(\wdp_a-\wdp*_b)t}{\rm d}t
 \\ \hphantom{ \int_{\Rset}K(\wdp*^-_a-\nu)\den*(\nu-\wdp_b)\,{\rm d}\nu}
 {}= K(\wdp*_a-\wdp_b-{\rm i})-K(\wdp_a-\wdp*_b),
 \\
 \int_{\Rset}K(\wdp*^-_a-\nu)\den*(\nu-\wdp*_b)\,{\rm d}\nu=0.
\end{gather*}

\subsection*{Acknowledgements}
The work of N.K.~was partially supported by the EIPHI Graduate School (contract ANR-17-EURE-0002). N.K.~would like to thank LPTMS laboratory (Universit\'e Paris-Saclay) and LPTHE laboratory (Sorbonne Universit\'e) for hospitality.
The work of G.K.~was supported by Carnot Pasteur doctoral school of the UBFC. We would like to thank the referees for several useful comments and suggestions.

\pdfbookmark[1]{References}{ref}
\LastPageEnding

\end{document}